\documentclass[aps,prd,superscriptaddress,nofootinbib,10pt,english,tightenlines,eqsecnum,floats,showpacs,,showkeys]{revtex4}
\usepackage{amstext,amssymb,amsfonts,amsmath}
\usepackage{fontenc}
\usepackage{textcomp}
\usepackage{mathrsfs}

\usepackage{color}
\usepackage{graphicx}
\usepackage{hyperref}

\newcommand{\f}[2]{\frac{#1}{#2}}
\newcommand{\scri}{{\mathscr I}}
\def\nn{\nonumber\\}
\def\be{\begin{equation}}
\def\ee{\end{equation}}
\def\bea{\begin{eqnarray}}
\def\eea{\end{eqnarray}}

\def\bwt{\begin{widetext}}
\def\ewt{\end{widetext}}

\begin{document}

\title{Trapped surfaces and nature of singularities in Lyra\rq{}s geometry}

\author{Amir Hadi Ziaie}\email{ah_ziaie@sbu.ac.ir}
\affiliation{Department of Physics, Shahid Beheshti University, G. C., Evin, Tehran, 19839, Iran}
\author{Arash Ranjbar}\email{a_ranjbar@sbu.ac.ir}
\affiliation{Department of Physics, Shahid Beheshti University, G. C., Evin, Tehran, 19839, Iran}
\author{Hamid Reza Sepangi}\email{hr-sepangi@sbu.ac.ir}
\affiliation{Department of Physics, Shahid Beheshti University, G. C., Evin, Tehran, 19839, Iran}

\date{\today}
\preprint{hep-th/yymmnnn}

\keywords{Gravitational collapse, naked singularities, Trapped surfaces, Lyra geometry}
\pacs{04.70.Bw, 04.20.Dw, 04.50.Kd}
\begin{abstract}
Motivated by the geometrical interpretation of Brans-Dicke (BD) scalar field which may also act as a torsion potential in Lyra geometry, we study the 
effects of spacetime torsion on the dynamics of a collapsing massive star. Taking the interior spacetime as the FLRW metric and the matter content as spherically symmetric, homogeneous perfect fluid with the equation of state (EoS) $p=w\rho$, we show that the collapse ends in a spacetime singularity which is of 
the strong curvature type in the sense of Tipler. Whether the trapped surfaces form during the dynamical evolution of the collapse depends on the torsion parameter, related to the BD coupling parameter, and the EoS subject to the conditions on physical reasonableness of the collapse configuration. Hence, the space of torsion and EoS parameters is divided into two portions, one for which the collapse process leads to the formation of apparent horizon and the other for which the apparent horizon is failed to form in the interior region. The nature of the singularity is examined from the exterior perspective, by searching for the existence of radial null geodesics reaching the faraway observers. Moreover, it is found that the effects of a dynamical torsion can be transferred to the outside region of the collapsing star, making the exterior region dynamic.

\end{abstract}

\maketitle
\section{Introduction}
An important and open issue in relativistic astrophysics is the final fate of a massive star collapsing under its own gravitational attraction. When a star with a mass  many times the Solar mass exhausts its nuclear fuel, the gravitational force overcomes the hydrostatic repulsive pressure and it starts to shrink and collapse continually
without ever reaching a final equilibrium state such as a white dwarf or a neutron star. Under physically reasonable circumstances, the singularity theorems then predict that the spacetimes describing the solutions to Einstein\rq{}s field
equations will inevitably admit singularities where densities and spacetime curvatures become infinitely large and blow up \cite{HE}.  However, these theorems predict only the existence of singularities and give no information on their nature. Thus, it is crucial to figure out whether, as hypothesized by cosmic censorship conjecture (CCC) \cite{CCCP}, the resulting singularity is necessarily covered by a spacetime event horizon (black hole) or can be observed by an external observer (naked singularity). The latter should, in principle, provide a laboratory for detecting the effects of quantum gravity near the extreme regions with ultra small length scales comparable to the Planck length. Despite lack of any mathematically rigorous formulation describing the cosmic censorship conjecture, several gravitational collapse models have been studied over the past years which exhibit the occurrence of naked singularities as a
possible outcome of a collapsing event. The models which have been investigated so far include a wide variety of field sources such as scalar fields \cite{SF}, perfect fluids \cite{PF} and imperfect fluids \cite{IPF}.
The problem can also be addressed in alternative theories of gravity such as $f(R)$ \cite{Khodam},  Lovelock \cite{Love} and Gauss-Bonnet gravity \cite{GBG}. 
The key feature pertaining to the visibility of a spacetime singularity to an external observer is the
causal structure of the spacetime at late stages of the collapse scenario that decides the possible
emergence or otherwise of non-spacelike trajectories from the resulting singularity. Therefore, if the
collapse dynamics proceeds in such
a way that the horizons develop much before the formation of the singularity, then extreme-dense regions
are causally disconnected from external observers and a black hole forms as the collapse end product.
On the other hand, if such horizons are delayed or failed to form during the collapse process, as governed
by the internal dynamics of the collapsing body, then a naked singularity is born which is nothing but
an ultra-strong gravity region communicating physical effects to the outside universe.

While mathematically permitted gravitational collapse models respect physical conditions such as
validity of the energy conditions and regularity of the initial data, a few of such models may
be applicable to a realistic star. Therefore, further physical features which affect the dynamics of a collapsing star is needed to be examined. Among these features, the inhomogeneities and related shearing effects have been vastly studied in the literature \cite{shear}. It has been shown that these effects could delay the formation of the apparent horizon, allowing regions of extremely strong gravitational fields to become causally connected to an outside observer.
Efforts directed towards the gravitational collapse of a massive star have mostly focused on the torsion-less spacetimes, while spacetimes with torsion have attracted less attention. In the presence of torsion, the connection could be generally asymmetric where its anti-symmetric part determines the torsion. Therefore, the geodesic equation is affected by introducing torsion. While the time-like particles with non-zero spin can directly interact with spacetime torsion, the spinless particles and photons which determine the causal structure of the spacetime interact indirectly through geometry\footnote{Particles with non-zero spin  move on the autoparallel curves over which a vector field is parallel transported via the affine connections. Photons and  other spinless particles follow the curves described by extremizing the line element with respect to the metric, i.e. the paths over which a vector field is parallel transported through the Christoffel symbol \cite{Hehl}.}.
Intuitively, one may think that the existence of spacetime torsion may modify the background spacetime and as a consequence affects the dynamics of the collapse in contrast to the case where torsion is absent.
Therefore, it may be of considerable interest to investigate the process of gravitationally collapsing massive objects when the effects of spacetime torsion are taken into account.

Shortly after Einstein\rq{}s theory of general relativity (GR), Weyl  introduced a generalization to the Riemannian geometry in order to unify gravitation and electromagnetism. However, this theory was never studied seriously due to its non-metricity. To cure this defect, a generalization to this theory was suggested by Lyra based on introducing a gauge function into the structure-less manifold \cite{Lyra}. In contrast to Weyl\rq{}s geometry which is torsion free but not metric compatible, Lyra\rq{}s connection is metric-preserving but not torsion free. Lyra\rq{}s action described in section \ref{review} is striking in the sense that the coupling constants are dimensionless. It conforms to the idea that a fundamental theory has no dimensional coupling and thus is scale invariant.
In Einstein-Cartan Theory, which is the simplest generalization of GR to include the intrinsic angular momentum (spin) of matter, torsion is not a dynamical object in the sense that its equation is purely algebraic and vanishes when the matter fields are absent \cite{VDSabbata-Gas}. On the other hand, in the scalar-tensor theory of Brans-Dicke (BD) \cite{BDT}, the tensor field alone is geometrized (it is specified by the metric tensor of Riemannian geometry)  and the scalar field remains alien to  geometry in the sense that the pure geometric essence of GR is not shared by the scalar-tensor theories.  Lyra's geometry is more in line with
Einstein's principle of geometerization, since both the scalar and tensor fields have intrinsic geometrical significance. From this viewpoint the BD theory can be obtained from pure geometry if one takes the fundamental
spacetime to be the Lyra manifold so that the BD scalar field casts either as the scale factor of
Lyra geometry or as the scalar torsion potential \cite{Soleng, BDLyra, KADUNN}.  In contrast to the BD theory where the coupling parameter is introduced in an {\it adhoc} fashion, when we re-write Lyra\rq{}s action in terms of the BD action, the origin of the coupling parameter appearing in the theory which measures the strength of the torsion is seen as rooted in the structure of the spacetime so that different manifolds will be distinguished through different connections.

Over the past decades, cosmological settings based on Lyra\rq{}s geometry have been broadly investigated.
In \cite{Sen-Ne}, it is shown that the redshift of spectral lines from extragalactic nebulae arises
as a result of an intrinsic geometrical essence of the model describing the universe. Several efforts have been made to clarify the possible role of Lyra geometry in a variety of contexts. Among them we quote, topological defects due to symmetry breaking in the early universe \cite{TOPO}, higher dimensional cosmological models \cite{HDL}, dark energy models \cite{DEL}, the study of spinorial field \cite{SPi}, massless DKP field \cite{MDKP}, scalar and vector massive fields \cite{SVMF} and semi-classical gravitational effects around global monopole \cite{SGGML}. Cosmological scenarios where the Big-Bang singularity is avoided have been found in \cite{LBB} (see \cite{2Singh} for a detailed review on cosmological settings in Lyra geometry) and black hole solutions in a Lyra background have been reported in \cite{LBH}. However, beside the cosmological scenarios where Lyra geometry has received a great deal of attention, to our knowledge, the process of gravitational collapse of a massive body has not yet been studied within this framework. While the setting presented herein may be regarded as a toy model, the question of what are the possible effects of a dynamical torsion on the final fate of a collapse scenario could be well motivated.

The aim of this paper is concerned with constructing a class of collapse models where the spacetime torsion propagates through a scalar potential and affects the collapse dynamics. We categorize the solutions based on the torsion parameter and EoS, subject to physical reasonableness of the collapse scenario into two classes: For the first class, we show that the footprint of the torsion can be traced as a frictional factor in the rate of change of the speed of collapse which increases the collapsing time interval (by reducing the speed of the collapse) and eventually retards the formation of the apparent horizon. However, for these class of solutions the apparent horizon would always form to cover the singularity. For the second class where the apparent horizon is failed to form, we find that the more we deviate from GR, the more the tendency of the solutions to the apparent horizon formation. The organization of the paper is as follows. After having a glance at  Lyra's geometry in section \ref{review}, we find the field equations and related solutions for a spherically symmetric homogeneous perfect fluid, benefiting from its connection to the BD theory. In section \ref{GC} we first present a quick summary on the formation or otherwise of trapped surfaces during a collapse scenario in GR, using the notion of Misner-Sharp gravitational energy. We then proceed by investigating the causal structure of the spacetime in BD theory and how the torsion could affect the collapse dynamics in relation to the variation of the torsion parameter and EoS. At the end of this section we deal with the issue of curvature strength of the singularity. In section \ref{ExtS} the exterior solution is presented and finally conclusions are drawn in the last section.

\section{Lyra geometry: Connection with Brans-Dicke theory}\label{review}
In Lyra geometry unlike the Riemannian geometry, where to each point on an $n$-manifold is assigned locally an $n$-tuple set of coordinates using a chart in each local neighborhood, the manifold is also endowed with a smooth scalar field $\psi$ at each point of $M$, called the gauge function, having the dimension of an inverse length so that the local coordinate system $x^{\mu}$ and the gauge function $\psi$ together form a {\it reference system}
\be
S=(\psi,x^{\mu}),\qquad \mu=1,...,n.
\ee
In fact, the gauge function plays the role of a scaling factor and scales the infinitesimal displacements
along a curve. Taking the preservation of the metricity condition in Lyra geometry as granted, one may find the form of the uniquely determined connection \cite{Soleng,Sen} as
\be\label{connection}
\Gamma^{\mu}\,_{\alpha\beta}=\frac{1}{\psi}\{^\mu\,_{\alpha\beta}\}+\frac{s+1}{\psi^2}(\delta^{\mu}
\,_{\beta}\psi_{,\alpha}-g_{\alpha\beta}\psi^{,\mu}),
\ee
where $s$ is a constant and reveals the properties of torsion in Lyra geometry which is defined by
\be\label{tor}
T^{\nu}\,_{\alpha\mu}=\frac{s}{\psi^2}(\delta^{\nu}\,_{\mu}
~\psi_{,\alpha}-\delta^{\nu}\,_{\alpha}~\psi_{,\mu}).
\ee
There is a close relation between Lyra geometry and Brans-Dicke theory in the sense that in the case of zero spin the field equations of both theories are equivalent \cite{Soleng}. To see this one can look at the action of the gravitational
theory based on Lyra\rq{}s manifold. The volume element
can be written in the form (from now on we proceed in $n=4 $ spacetime dimension)
\be
\eta=\psi^4\,\sqrt{-g}~d^4x.
\ee
The Lyra\rq{}s gravitational action is constructed, in analogy to the Riemannian case,
as
\be
{\mathcal S}=\int_{\mathcal M}K~\eta +{\mathcal S_m},
\ee
where ${\mathcal S_m}$ stands for the action for matter fields and the Lyra curvature scalar $K$ is written in the form of
\be\label{lyra-scalar}
K=\frac{1}{\psi^2}R+\frac{2(s+1)}{\psi^3}(1-n)\square\psi+\frac{1}{\psi^4}
\Bigg[(s+1)^2(3n-n^2-2)-2(s+1)(1-n)\Bigg]\psi_{,\alpha}\psi^{,\alpha},
\ee
where $\square$ is the Riemann-Christoffel d\rq{}Alembert operator.
Invoking  equation (\ref{lyra-scalar}) and eliminating the total divergence terms, we obtain the BD action
\be
{\mathcal S}=\int d^4x~\sqrt{-g}~\bigg( \phi R -\frac{\omega}{\phi}g^{\mu\nu}\partial_{\mu}\phi~\partial_{\nu}\phi\bigg)+{\mathcal S_m},
\ee
where we have set $\psi^2=\phi$ and $\omega=\frac{3}{2}(s^2-1)$ is the BD coupling parameter. Extremizing the above action one
gets the modified BD field equations as
\bea
G_{\mu\nu}&=&\frac{8\pi}{\phi}T^m_{\mu\nu}+\frac{\omega}{\phi^2}\bigg[\partial_{\mu}\phi~\partial_{\nu}\phi-\frac{1}{2}g_{\mu\nu}\partial_{\alpha}\phi~\partial^{\alpha}\phi\bigg]+\frac{1}{\phi}\bigg[\phi_{,\mu;\nu}-g_{\mu\nu}~\square\phi\bigg],\label{Einstein}\\
\square\phi&=&\frac{8\pi}{3s^2} T^m,\label{evolution}
\eea
where $T^m$ stands for the trace of the energy-momentum tensor of matter fields of spinless particles  and we have set $G=c=1$. In the presence of spin-torsion interactions however, an additional term from spin density comes into  play through the evolution equation for torsion potential \cite{Soleng}. We note that the condition $s^2\geq 0$ restricts the BD coupling parameter to be greater than or equal to $-3/2$ . For $s=0$ the spacetime torsion vanishes and the Lyra geometry reduces to a (local) conformally invariant Riemanian geometry (see \cite{Soleng} for more details). Regarding the relation of Lyra geometry and BD theory, this case corresponds to special value of the BD coupling parameter for which the Cauchy problem is ill-posed \cite{-3/2}, hence we exclude it.

Next, within the framework introduced above, we construct a class of continual collapse models for a spherically symmetric homogeneous matter distribution in such a way that the collapse evolution terminates in a spacetime singularity. We  parameterize the interior line element of the collapsing body, in comoving coordinates, as
\begin{equation}\label{metric}
ds^2=-d\tau^2+a(\tau)^2dr^2+R(\tau,r)^2d\Omega^2,
\end{equation}
where $R(\tau,r)=ra(\tau)$ is the physical area radius of the collapsing volume and $\tau$ is the comoving time.
We take the matter energy-momentum tensor to be that of a perfect fluid with the barotropic EoS $p=w\rho$
\be
T^{^m}\!^{\mu}\,_{\nu}={\rm diag}(-\rho_{_m}, p_{_m}, p_{_m}, p_{_m}),
\ee
where $\rho_{_m}$ and $p_{_m}$ are the energy density and pressure.

Inserting  metric (\ref{metric}) into the field equations (\ref{Einstein}) and (\ref{evolution}) one finds three independent equations
\bea
3H^2&&=\frac{3}{4}(s^2-1)\left(\frac{\dot{\phi}}{\phi}\right)^2-3H\frac{\dot{\phi}}{\phi}+\frac{8\pi\rho_{_{m}}}{\phi}=8\pi\rho_{{\rm eff}},\label{00comp}\\
-2\dot{H}-3H^2&&=2H\frac{\dot{\phi}}{\phi}+\frac{\ddot{\phi}}{\phi}+\frac{3}{4}(s^2-1)\left(\frac{\dot{\phi}}{\phi}\right)^2+\frac{8\pi p_{_m}}{\phi}=8\pi p_{{\rm eff}},\label{11comp}\\
\frac{\ddot{\phi}}{\phi}&&=\frac{8\pi}{3s^2\phi}(\rho_m-3p_m)-3H\frac{\dot{\phi}}{\phi}\label{evol},
\eea
where $H(\tau)=\f{\dot{R}(\tau,r)}{R(\tau,r)}=\f{\dot{a}(\tau)}{a(\tau)}$ .
Substituting for $\f{\ddot{\phi}}{\phi}$ from (\ref{evol}) into equation (\ref{11comp}) and after
a bit of rearrangement, we obtain
\bea
3H^2&=&\frac{3}{4}(s^2-1)a^2H^2\left(\f{\phi_{,a}}{\phi}\right)^2-3aH^2\f{\phi_{,a}}{\phi}+\f{8\pi\rho_{i_{m}}}{\phi}a^{-3(1+w)}\label{re00},\\
-2aHH_{,a}-3H^2&=&-aH^2\f{\phi_{,a}}{\phi}+\f{3}{4}\left(s^2-1\right)a^2H^2\left(\f{\phi_{,a}}{\phi}\right)^2+8\pi\rho_{i_{m}}\f{1+3w(s^2-1)}{3s^2\phi}a^{-3(1+w)}\label{re11},
\eea
where the conservation of ordinary matter energy-momentum tensor
${\dot{\rho}}_{_m}+3H(\rho_{_m}+p_{_m})=0$ has been considered and $\rho_{i_{m}}$ is the initial
profile of the energy density at initial hypersurface from which the collapse commences.
Solving the above set of differential equations together with the evolution equation for the BD scalar field (\ref{evol}) we get
\bea
H(a(\tau))=\alpha a(\tau)^{\gamma},~~~~\phi(a(\tau))=\beta a(\tau)^{\delta},\label{H}
\eea
where
\be\label{Coeff}
\alpha=\sqrt{\frac{8 \pi  \rho_{i_{m}}}{3 \beta s^2}}\frac{\left(1-3 s^2 (1-w)-3 w\right)}{ \left(9 s^2 (w-1)^2-(3 w-1)^2\right)^{\f{1}{2}}},~~\gamma=\frac{9 w^2-9 s^2 \left(w^2-1\right)-1}{6 s^2 (w-1)-6 w+2},~\,\,\delta=\frac{2 (3 w-1)}{1-3 w-3 s^2 (1-w)},~\,\,s\neq0,
\ee
and $\beta$ is a constant. Since we are concerned with a collapsing configuration, the first part of
equation (\ref{H}) with $\alpha<0$, which measures the rate of decrease of the area radius, will be used throughout this paper. It is then easy to solve for the scale factor as a function of the comoving time as
\be\label{scf}
a(\tau)=\Bigg[a_i^{-\gamma}+\alpha \gamma(\tau_i-\tau)\Bigg]^{-\f{1}{\gamma}},
\ee
where $a_i$ and $\tau_i$ are the initial values of the scale factor and comoving time
respectively. In order that the scale factor vanishes at a finite amount of comoving time we require $\gamma<0$ so that the singularity time can be obtained as
\be
\tau_s=\tau_i+\f{a_i^{-\gamma}}{\alpha \gamma}.
\ee

\section{Gravitational collapse}\label{GC}
\subsection{A short review}
In this subsection we first give a brief review on active gravitational energy and then proceed by
examining our solution in a continual spherically symmetric  gravitational collapse. Basically, it is the structure of trapped surfaces during the collapse procedure that decides the visibility or otherwise
of the spacetime singularity. These surfaces are defined as compact two-dimensional (smooth)
space-like surfaces such that both families of ingoing and outgoing null geodesics normal
to them necessarily converge \cite{Frolov}. A point is considered as being in a trapped region if
there exists a trapped surface surrounding it. A trapped region is defined as
the union of all trapped surfaces. The interior spacetime (\ref{metric}) can be split into
the surface of a 2-sphere and a two dimensional hyper-surface normal to the 2-sphere as \cite{Hayward}
\begin{equation}\label{hmunu}
ds^2=h_{\mu\nu}dx^{\mu}dx^{\nu}+R(\tau,r)^2d\Omega^2,~~~~~h_{\mu \nu}={\rm diag}\left[-1,a(\tau)^2\right].
\end{equation}
Introducing the null coordinates
\begin{equation}\label{doublenull}
d\zeta^{+}=-\frac{1}{\sqrt{2}}\left[d\tau-a(\tau)dr\right],~~~~d\zeta^{-}=-\frac{1}{\sqrt{2}}\left[d\tau+a(\tau)dr\right],
\end{equation}
the above metric can be re-written in a double-null form
\begin{equation}\label{metricdnull}
ds^2=-2d\zeta^{+}d\zeta^{-}+R(\tau,r)^2d\Omega^2.
\end{equation}
The radial null geodesics are given by the condition $ds^2=0$. Thus there exist two kinds of
future-directed null geodesics corresponding to $\zeta^{+}=constant$ and $\zeta^{-}=constant$, the
expansions of which are given by
\begin{equation}\label{expansion}
\theta_{\pm}=\frac{2}{R(\tau,r)}\partial_{\pm}R(\tau,r),
\end{equation}
with the partial derivatives taken along the null coordinates
\begin{equation}\label{p+p_}
\partial_{+}=\frac{\partial}{\partial\zeta^{+}}=\frac{1}{\sqrt{2}}\left[\partial_{\tau}+\frac{\partial_{r}}{a(\tau)}\right],~~~\partial_{-}=\frac{\partial}{\partial\zeta^{-}}=\frac{1}{\sqrt{2}}\left[\partial_{\tau}-\frac{\partial_{r}}{a(\tau)}\right].
\end{equation}
The expansion parameter measures whether the congruence of null rays normal to a sphere is
diverging $(\theta_{\pm}>0)$ or converging $(\theta_{\pm}<0)$, in other words, the area
radius along the light rays is increasing or decreasing, respectively. The spacetime is referred to
as trapped, untrapped and marginally trapped if \footnote{Note that, the objects
$\theta_{+}$ and $\theta_{-}$ are not geometrically invariant since the null coordinates
$\zeta^{+}$ and $\zeta^{-}$ can be arbitrarily rescaled as $\zeta^{+}\rightarrow f(\zeta^{+})$,
$\zeta^{-}\rightarrow g(\zeta^{-})$. An invariant object is the combination $\theta_{+}\theta_{-}$.}
\begin{equation}\label{sp}
\theta_{+}\theta_{-}>0,~~~~ \theta_{+}\theta_{-}<0,~~~~\theta_{+}\theta_{-}=0,
\end{equation}
respectively, where the third class characterizes the outermost boundary of the trapped region, the apparent horizon.
From equation (\ref{expansion}) one can easily check that
$h^{\mu\nu}\partial_{\mu}R(\tau,r)\,\partial_{\nu}R(\tau,r)=-R(\tau,r)^2\theta_{+}\theta_{-}/2$, thus the apparent horizon
curve is simply given by the condition $\dot{R}(\tau,r)^2=1$. In other words, the vector $\partial^{\mu}R$
is space-like (time-like) in untrapped (trapped) regions and null on the apparent horizon. The Misner-Sharp (MS)
gravitational energy in GR is defined as \cite{Misner-Sharp} (see also \cite{JAP} for more details)
\begin{equation}\label{MS}
E(\tau,r)=\frac{R(\tau,r)}{2}\left[1-h^{\mu\nu}\partial_{\mu}R(\tau,r)\,\partial_{\nu}R(\tau,r)\right]=\frac{R(\tau,r)}{2}\left[1+
\frac{R(\tau,r)^2}{2}\theta_{+}\theta_{-}\right].
\end{equation}
From the above definition, it is the ratio $2E(\tau,r)/R(\tau,r)$ that governs the formation of trapped surfaces during the dynamical evolution of the collapse scenario such that, if the condition $2E(\tau,r)<R(\tau,r)$ holds
then no trapping of light happens and the apparent horizon (which is the outermost boundary of trapped surfaces), characterized by $2E(\tau_{{\rm AH}},r)=R(\tau_{{\rm AH}},r)$,
is not formed early enough before the singularity formation or is totally avoided ($\tau_{s}<\tau_{{\rm AH}}$). In this situation the outgoing light rays may have a chance to escape toward a neighboring observer (locally naked singularity)
or an asymptotic one (globally naked singularity). For $2E(\tau,r)>R(\tau,r)$, trapped surfaces do form
throughout the  evolution of the collapse  leading to the convergence of both the ingoing and outgoing
families of null trajectories that have emerged from a point existing in the trapped region. Thus, the
apparent horizon forms early enough before the singularity formation produces a black-hole in the
spacetime ($\tau_{s}>\tau_{{\rm AH}}$).
\subsection{Causal structure of the spacetime and dynamics of the apparent horizon}\label{GMS}
\begin{figure}
\includegraphics[scale=0.7]{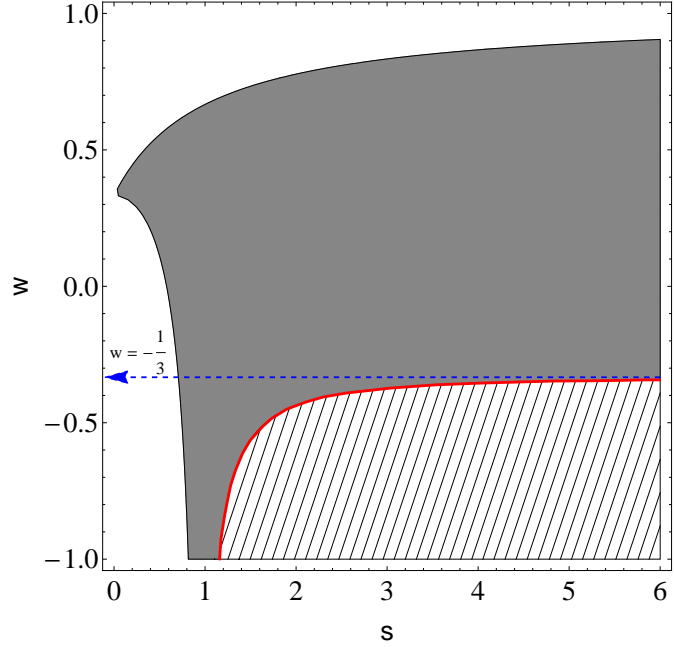}
\caption{The allowed region for trapped surface formation (gray zone) or their avoidance (shaded zone). These regions are plotted by requiring that $\alpha<0$ since the collapse rate must be negative, $\gamma<0$ in order that the scale factor vanishes at a finite amount of comoving time, $\delta>0$ so that there exist null trajectories terminating in the past at the singularity from the exterior viewpoint and the WEC be satisfied throughout the collapse setting. The red boundary separates these regions and asymptotes the line $w=-\f{1}{3}$ (the blue dotted arrow) for large values of the $s$ parameter.}\label{regionn}
\end{figure}

\begin{figure}
\begin{center}
\includegraphics[scale=0.55]{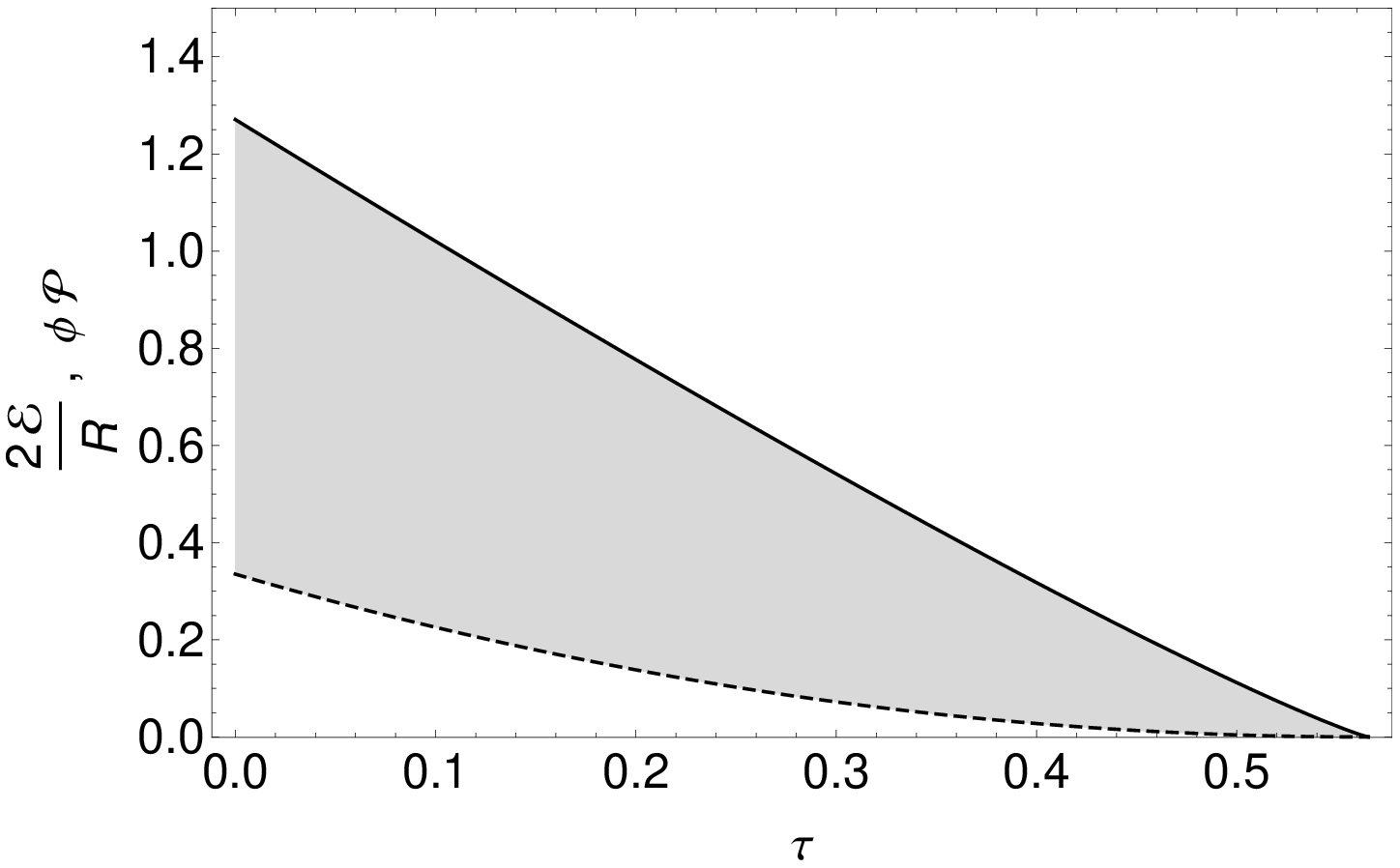}\hspace{4mm}
\includegraphics[scale=0.55]{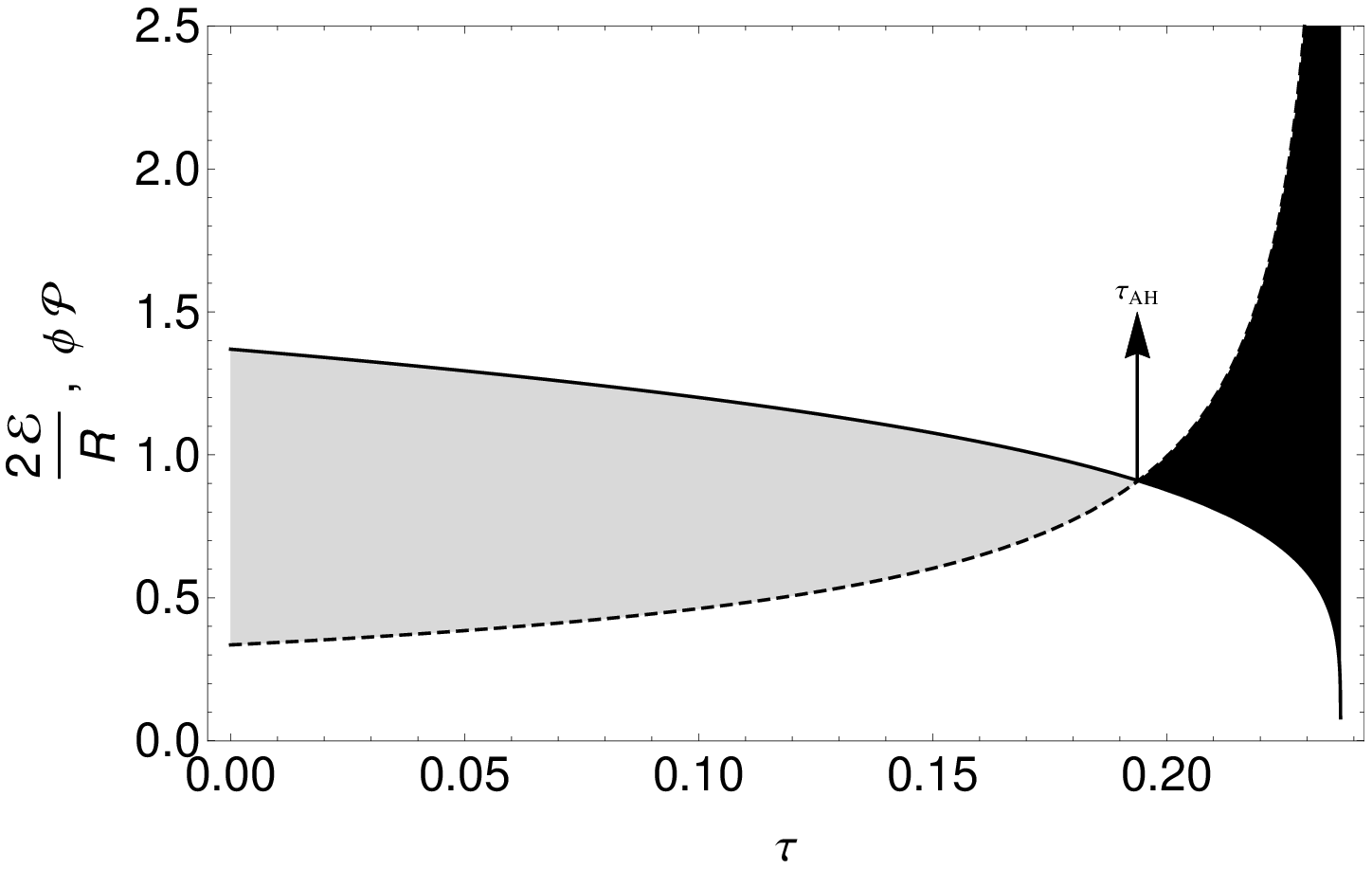}\vspace{4mm}
\includegraphics[scale=0.55]{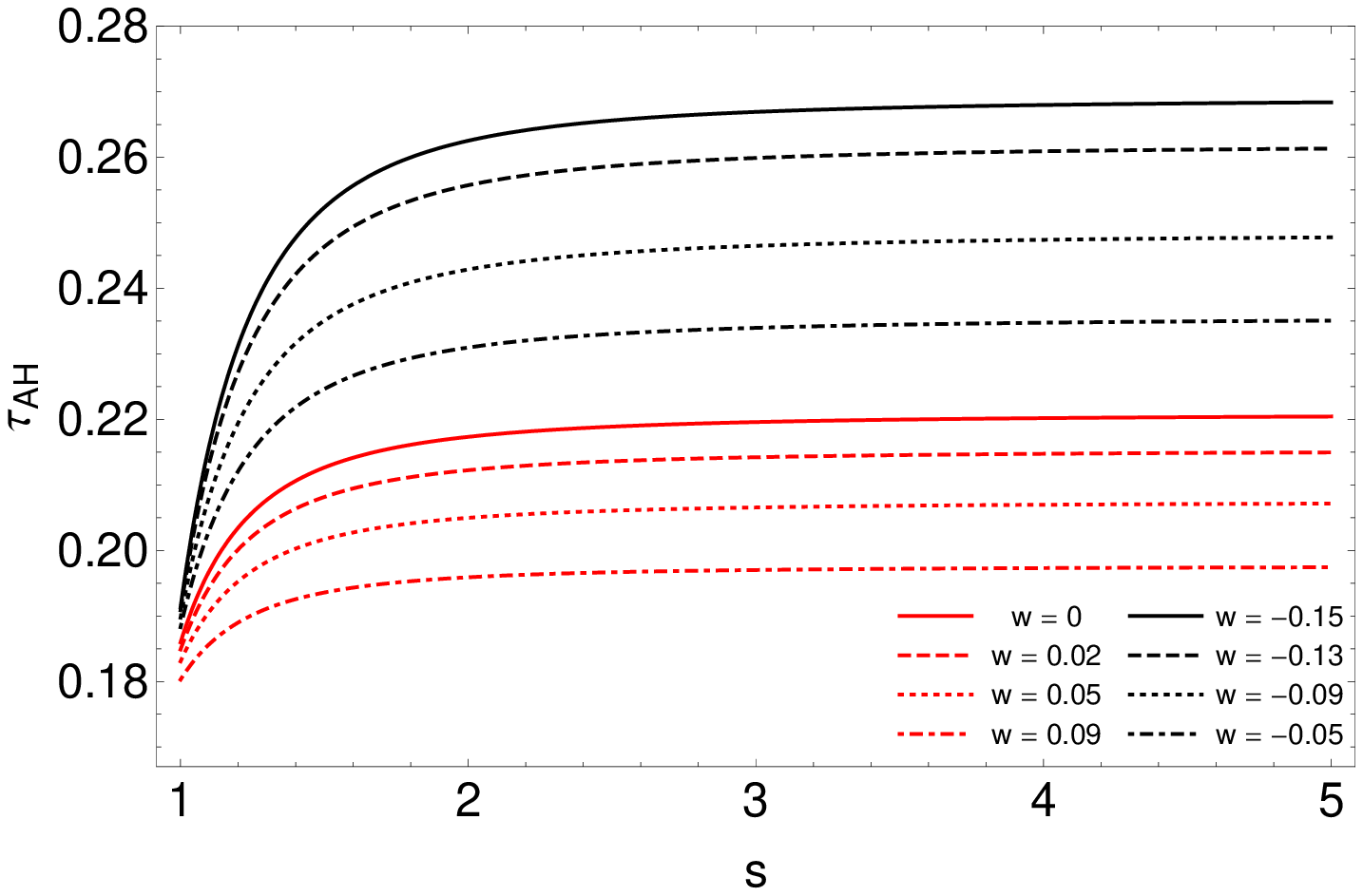}\hspace{4mm}
\includegraphics[scale=0.55]{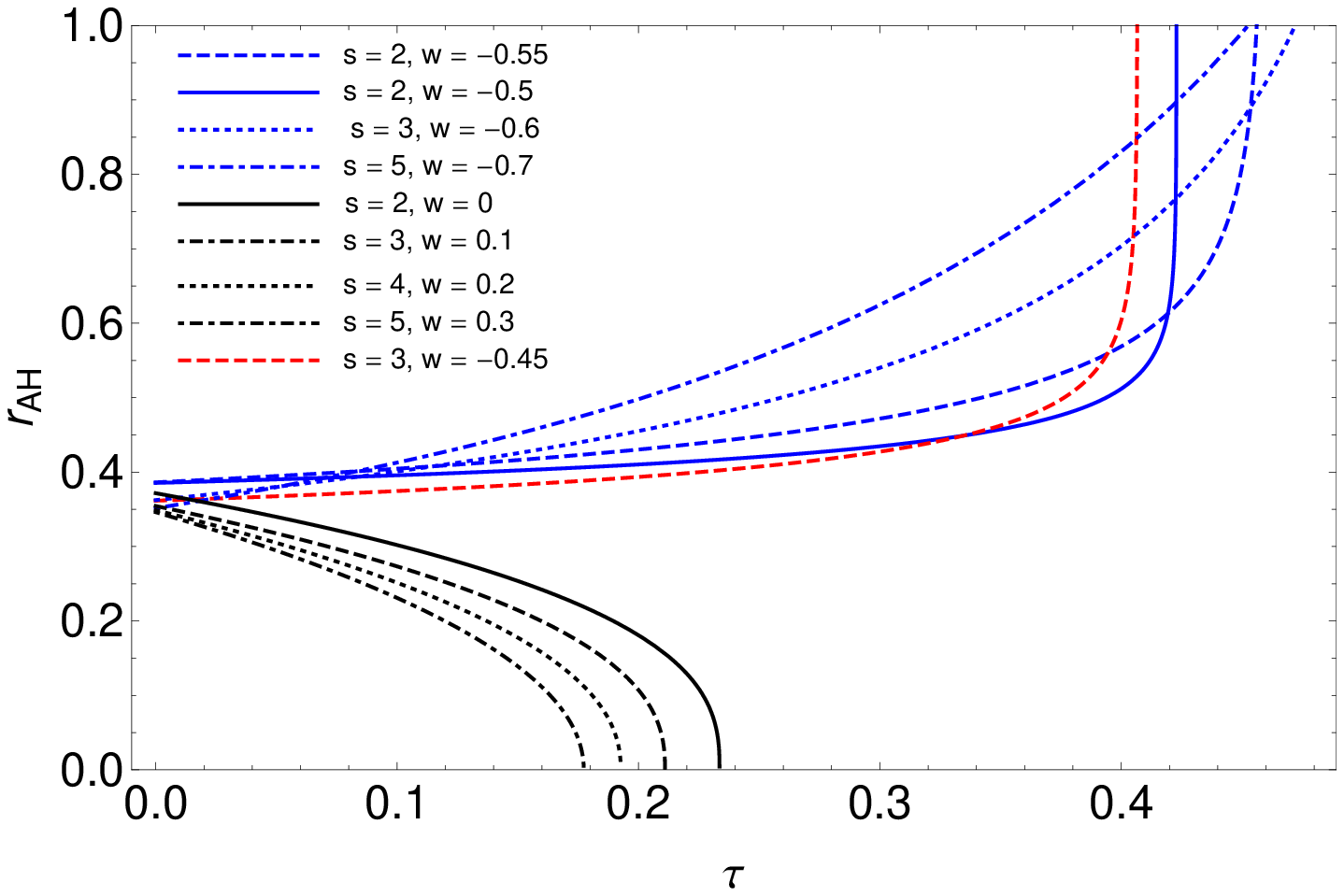}
\caption{Upper left panel: The behavior of $2{\mathcal E}/R$ (dashed curve) and $\phi\mathcal{P}$ (solid curve) as functions of $\tau$ for $\beta=0.8$, $\rho_{i_{m}}=1$, $a_i=1$ and $\tau_i=0$. The torsion and EoS parameters have been picked up from the shaded region of figure (\ref{regionn}) as $s=1.5$, $w=-0.8$. The ratio $2{\mathcal E}/R/\phi\mathcal{P}$ stays less then unity till the singularity time $\tau_s=0.565$. Here, use is made of the scaling freedom for the initial radial comoving coordinate to write $R(a(\tau_i),r)=r=0.2$. Note that we have to suitably utilize this scaling freedom so that the regularity condition is fulfilled.
Upper right panel: Apparent horizon formation for $\beta=1$, $\rho_{i_{m}}=1$, $a_i=1$ and $\tau_i=0$ at $\tau_{{\rm AH}}=0.193$. The torsion and EoS parameters have been picked up from the gray region of figure (\ref{regionn}) as $s=1.4$, $w=0$.
Lower left panel: Effects of variation of $s$ parameter on delay in the apparent horizon formation for $\beta=1$, $\rho_{i_{m}}=1$, $a_i=1$, $\tau_i=0$ and $r=0.2$. The numerical values of torsion and EoS parameters have been chosen from the gray region of figure (\ref{regionn}). Lower right panel: The behavior of apparent horizon curve for different values of torsion and EoS parameters.}\label{Fig2}
\end{center}
\end{figure}

\begin{figure}
\includegraphics[scale=0.52]{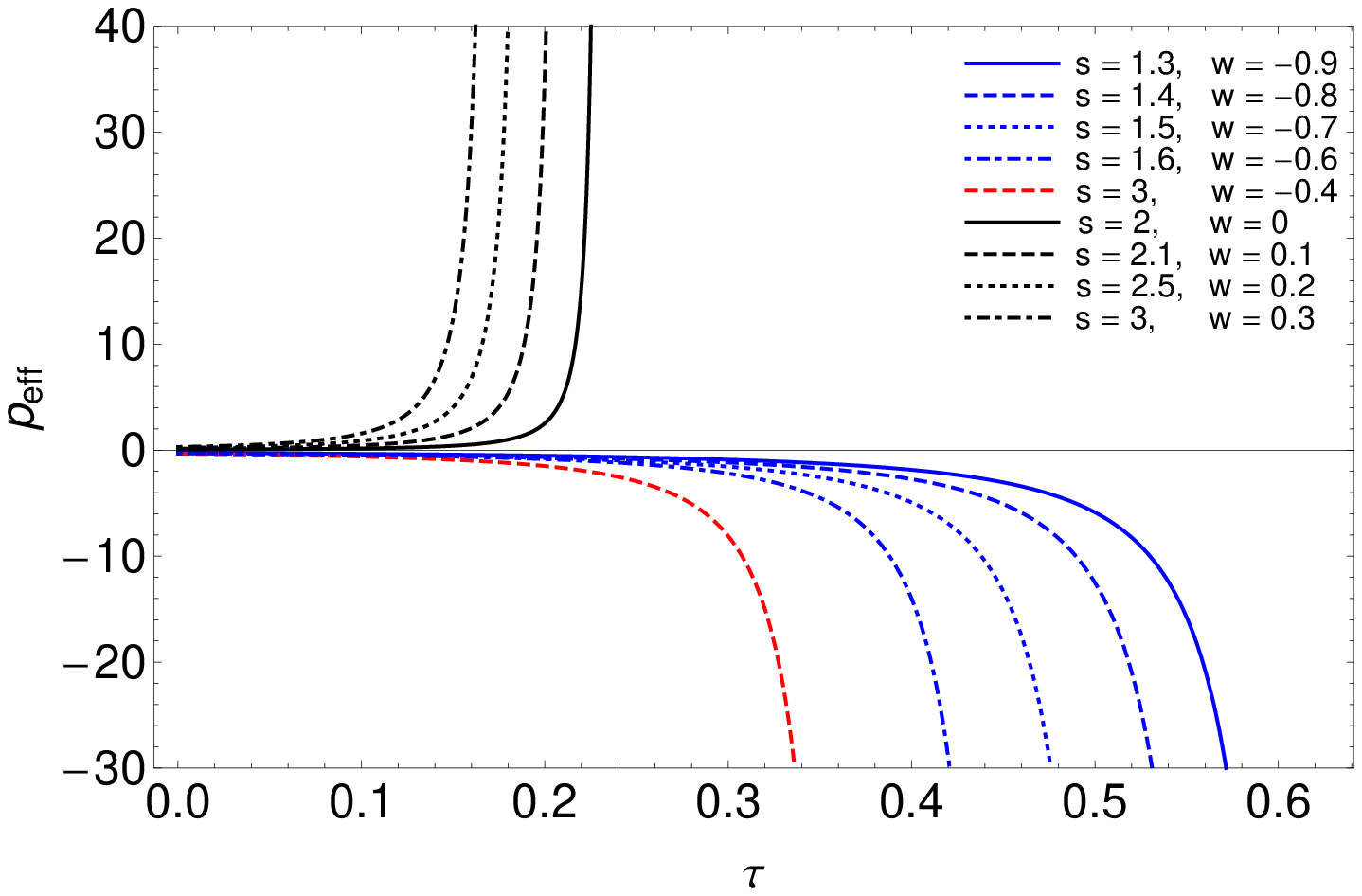}
\includegraphics[scale=0.5]{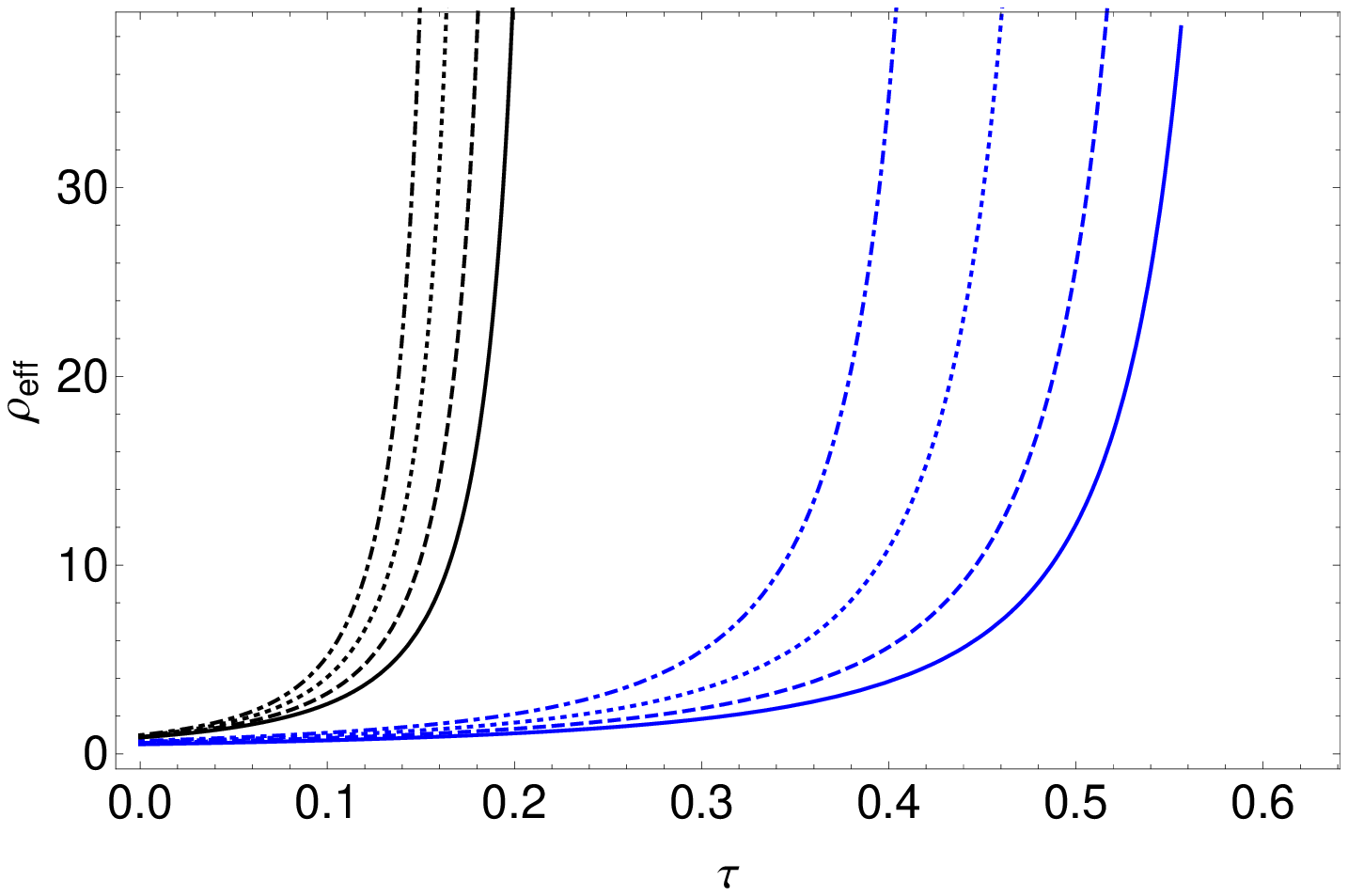}
\includegraphics[scale=0.5]{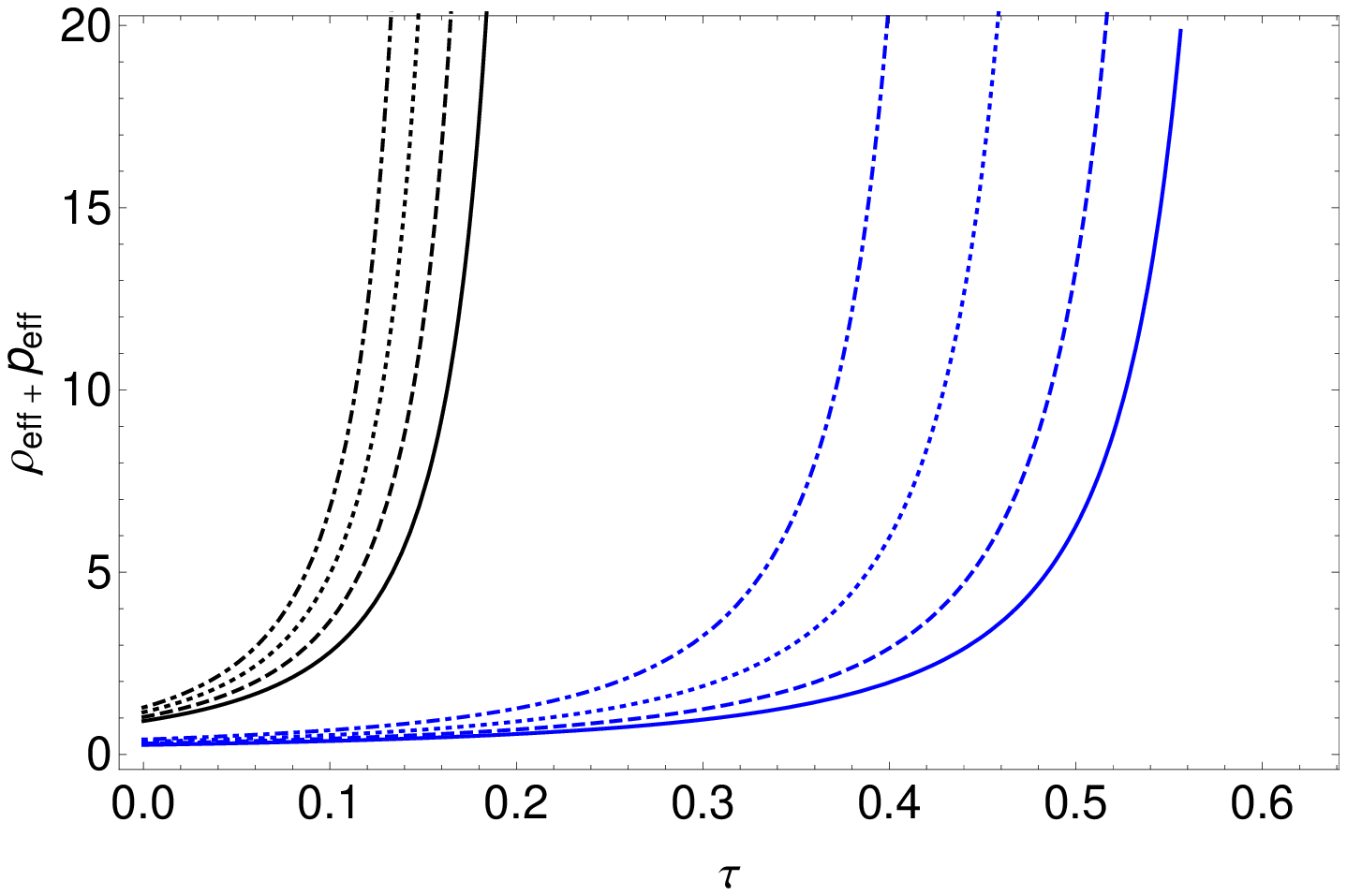}
\includegraphics[scale=0.55]{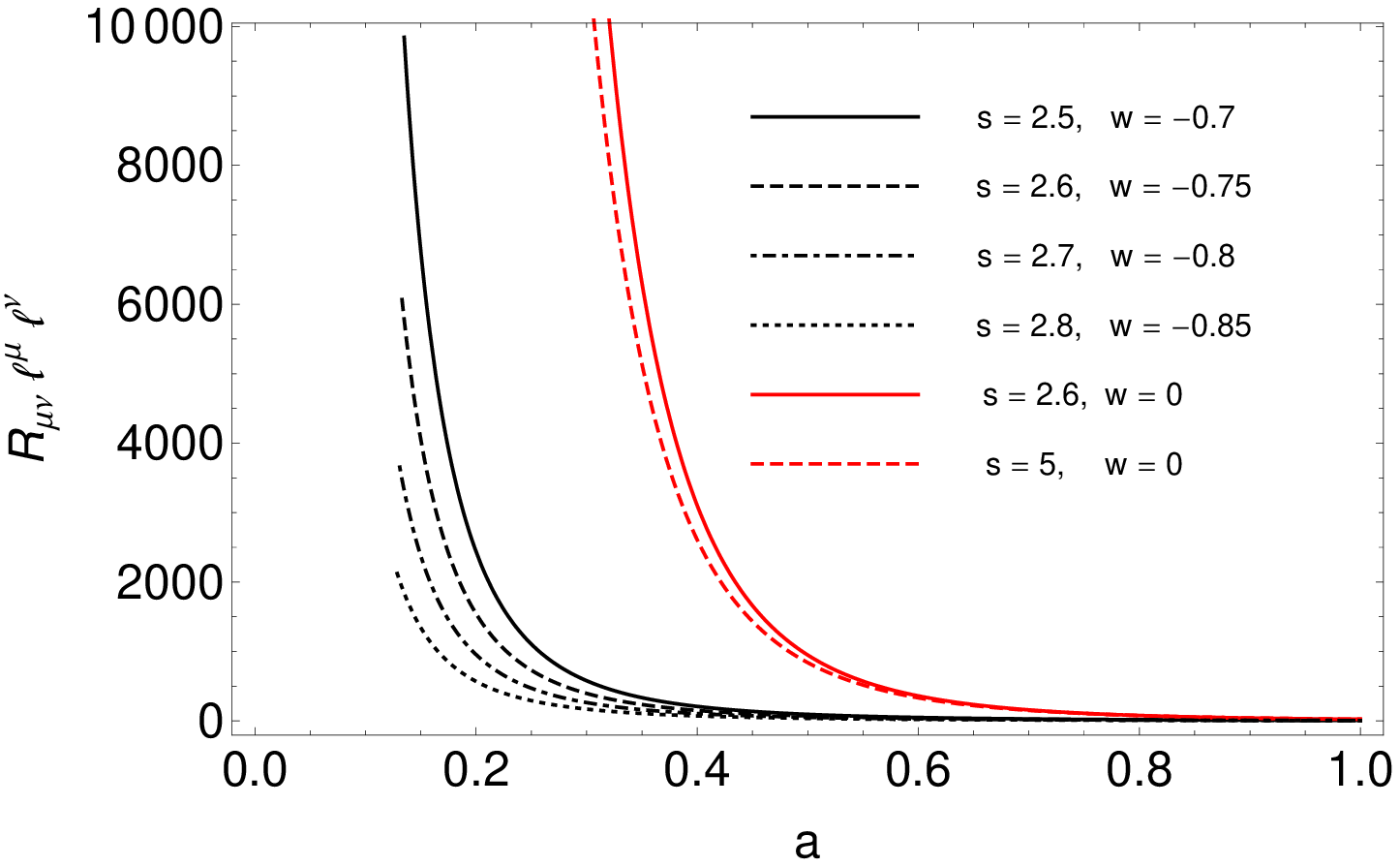}
\caption{Upper Left panel: The time behavior of effective pressure for different values of the torsion and EoS parameters. Upper right panel: The time behavior of the effective energy density for the same values of $s$ and $w$ parameters as the left figure. Lower left panel: The weak energy condition for the same values of $w$ and $s$ parameters as upper panels. We have set $\beta=1$ and the initial values of energy density, the scale factor and initial time as, $\rho_{i_{\rm m}}=1$, $a_i=1$ and $t_i=0$, respectively. Lower right panel: the behavior of null curvature condition for different values of torsion and EoS parameters.}\label{WECF}
\end{figure}

In the previous subsection we defined the MS energy that determines the causal structure of the spacetime in GR. Now consider a spacetime 
manifold on which a scalar field has been spread out so that this scalar field couples only indirectly to the matter fields but non-minimally interacts 
with the curvature through the coupling to the Ricci scalar. In this sense, one expects that the gravitational interaction is mediated not only by the usual 
metric tensor field of GR, but also by an additional scalar field. Hence, the causal structure of the manifold endowed with this scalar field gets altered 
in comparison to the case where the scalar field is absent. Therefore, we define the mass-like function \cite{GMSMAG} which in the case of GR
is reduced to the standard MS energy. In order to compute the mass-like function we follow the procedure suggested in \cite{cai} using the unified first 
law of thermodynamics. Upon using this method the field equations can be rewritten as
\be
d{\mathcal E}=A\Psi_{\mu}dx^{\mu}+WdV,
\ee
where $A=4\pi R(\tau,r)^2$ is the area of the sphere with physical radius $R(\tau,r)$,  $V=(4\pi/3) R(\tau,r)^3$
is its volume, $W$ is the work density  $W=-(1/2) h_{\mu\nu}T_{_m}^{\mu\nu}$ and $\Psi_{\mu}=T^{\nu}_{{_m}\mu}\partial_{\nu}R(\tau,r)+W\partial_{\mu} R(\tau,r)$ is the energy supply vector.
It is straightforward to show that
\be
d{\mathcal E}=A(\tau,r)d\tau+B(\tau,r)dr,
\ee
where
\bea
A(\tau,r)&=&r^3 a^3 H\bigg[\phi \f{\ddot{a}}{a}+\f{\phi}{2}H^2+\dot{\phi}H+\f{\ddot{\phi}}{2}+\f{3(s^2-1)}{8}\f{\dot{\phi}^2}{\phi}\bigg],\label{A}\\
B(\tau,r)&=&r^2 a^3\bigg[\f{3}{2}\phi H^2-\f{3(s^2-1)}{8}\f{\dot{\phi}^2}{\phi}+\f{3}{2}\dot{\phi}H\bigg].\label{B}
\eea
For $d{\mathcal E}$ to be a total differential the following integrability condition has to be satisfied
\be
\f{\partial A(\tau,r)}{\partial r}-\f{\partial B(\tau,r)}{\partial\tau}=0,
\ee
whence we get
\footnote{Since $d{\mathcal E}$ is a total differential
\be
{\mathcal E}(\tau,r)=\int B(\tau,r) dr +\int \big[ A(\tau,r)-\f{\partial}{\partial\tau}\int B(\tau,r) dr\big] d\tau,
\ee
is another solution for the mass-like function equivalent to equation (\ref{MS-int}).}
\be\label{MS-int}
{\mathcal E}(\tau,r)=\int A(\tau,r) d\tau +\int \bigg[ B(\tau,r)-\f{\partial}{\partial r}\int A(\tau,r) d\tau\bigg] dr.
\ee
It is worth noting that  equation (\ref{evol}) implies the satisfaction of the integrability condition.
Substituting the expressions (\ref{A}) and (\ref{B}) into the above equation and taking into account the fact that $A(\tau,r)$
and $B(\tau,r)$ are separable functions of $r$ and $\tau$, we arrive at the following equation for the mass-like function as
\be
{\mathcal E}(\tau,r)=\int B(\tau,r) dr= \f{r}{3} B(\tau,r).
\ee
Therefore one can find
\be\label{MS-energy}
{\mathcal E}(\tau,r)=\f{1}{2}R(\tau,r)^3\phi\bigg[H^2-\f{s^2-1}{4}\left(\f{\dot{\phi}}{\phi}\right)^2+\f{\dot{\phi}}{\phi}H\bigg].
\ee
Rewriting the above equation in terms of the invariant quantity $\theta_{+}\theta_{-}$, one has
\be\label{rewriteMS}
{\mathcal E}=\f{R\phi}{2}{\mathcal P}\left[1+\f{R^2}{2}\theta_{+}\theta_{-}\right],
\ee
where
\be
{\mathcal P}=\left[1+a\left(\f{\phi_{,a}}{\phi}\right)-\f{1}{4}a^2(s^2-1)\left(\f{\phi_{,a}}{\phi}\right)^2\right],
\ee
which, for the solutions obtained from the second part of (\ref{H}) reads ${\mathcal P}=8\pi\rho_{i_{m}}/3\beta\alpha^2$. Bearing in mind the conditions given in (\ref{sp}), the spacetime is trapped, untrapped or marginally
trapped if
\be\label{ratio-ps}
\f{2{\mathcal E}}{R}>\phi {\mathcal P},\quad \f{2{\mathcal E}}{R}<\phi {\mathcal P},\quad \f{2{\mathcal E}}{R}=\phi {\mathcal P},
\ee
respectively. We note that in case where $s\rightarrow \infty$ and the trace of energy-momentum tensor of matter is non-vanishing, the quantity ${\mathcal P}\rightarrow 1$, the BD scalar field becomes a constant and the above relations reduce to their GR counterparts\footnote{It should be noticed that in Lyra geometry for $s=0$ the autoparallel curves and geodesics coincide, though the connection is manifestly asymmetric that is because the coordinate frame is not chosen as the natural one. However, as it is clear from equation (\ref{tor}) the torsion vanishes for $s=0$. This case which corresponds to a (local) conformal Riemannian manifold with a local conformal weight $\psi(x)$ (or a Riemannian manifold which is endowed with a BD scalar field with a dimensionless coupling parameter being equal to $-3/2$) is excluded because of ill-defined Cauchy problem as it was already mentioned. It is important to notice that the GR limit of the BD theory is realized for $\omega \rightarrow \infty$ and $T^{m}\neq0$, as equation (\ref{evolution}) dictates, which results in $\psi(x)=\textrm{constant}$. In this case the Lyra manifold reduces to a conformal Riemannian manifold with a constant global conformal weight. In this sense, GR is a smooth limit of the Lyra geometry in the regime that $\omega \rightarrow \infty$.} (see \cite{MEHR} for a detailed discussion on the GR limit of the BD theory). Using equations (\ref{H}) and (\ref{MS-energy}) it is easy to find the following expressions as
\be\label{ratio}
\f{2{\mathcal E}(\tau,r)}{R(\tau,r)}=\f{8\pi r^2\rho_{i_{m}}}{3}\left[a_i^{-\gamma}+\alpha \gamma(\tau_i-\tau)\right]^{\f{1+3w}{\gamma}},~~~~ \phi(\tau)=\beta\Bigg[a_i^{-\gamma}+\alpha \gamma(\tau_i-\tau)\Bigg]^{-\f{\delta}{\gamma}}.
\ee
The solution (\ref{scf}) exhibits a curvature singularity, a point at which both Lyra-Kretschmann scalar (see appendix \ref{appA}) and the effective energy density (\ref{00comp}) diverge. Then, in order to determine whether such a singularity is hidden behind a horizon or not, one needs to investigate the behavior of trapped surfaces during the dynamical evolution of the collapse procedure. Firstly, we note that the regularity condition needs to be satisfied which states that there should not be any trapping of light at the initial hypersurface from which the collapse begins, $2{\mathcal E}/R|_{\tau_i}<\phi\mathcal{P}|_{\tau_i}$.
Eventually, if the second part of equation (\ref{ratio-ps}) holds throughout the collapse until the
singularity epoch $\tau_s$ is reached, trapped surface formation is avoided throughout the collapse
scenario and a naked singularity may form as the collapse end result. If there is a time
for which the first part of equation (\ref{ratio-ps}) holds then the trapped surfaces would form and the resulting singularity will be necessarily covered. Secondly, the sign of the pressure plays an important role in determining the final outcome of the collapse scenario. In a typical continual collapse scenario the absence of trapped surfaces is usually accompanied by negative pressure \cite{NEGP,COOP}. However, for the sake of physical reasonableness, it is required that the collapse process obeys the weak energy condition (WEC), i.e., $T_{ij}V^{i}V^{j}\geq0$ for all non-spacelike vectors $V^{i}$, that is, the energy density as measured by any local observer is non-negative (we note that the weak energy condition implies the null one). For the collapse setting considered here, the weak energy condition is equivalent to the following statements
\be\label{WEC}
\rho_{{\rm eff}}\geq0,~~~~\rho_{{\rm eff}}+p_{{\rm eff}}\geq0.
\ee
Figure (\ref{regionn}) shows the allowed regions in $(s,w)$ plane in the sense that in the collapse setting, the scale factor vanishes at a finite amount of time subject to the regularity and energy conditions (we require that $\alpha<0$ and $\delta>0$). For any point selected for the torsion parameter ($s$) and EoS ($w$) from the gray region, trapped surfaces do form during the collapse procedure. While for those chosen from the shaded region, the formation of trapped surfaces is avoided. The boundary that separates these two regions is highlighted by a red curve. This boundary coincides with the line $w=-\f{1}{3}$ as the $s$ parameter goes to infinity (as asymptotically given by the blue dotted arrow). Thus, for $s\rightarrow\infty$ or correspondingly for $\omega\rightarrow\infty$, this boundary separates the trapped and untrapped regions as simply given by $w>-\f{1}{3}$ and $w<-\f{1}{3}$, respectively. This is nothing but the GR limit of the model. We further note that the gray region enclosed between the red curve and $w<-\f{1}{3}$ contains a set of points in $(s,w)$ plane for which the collapse leads to black hole formation. While in the GR case, this region must have been the one for which trapped surfaces are avoided for all $-1<w<-\f{1}{3}$, the presence of Lyra torsion puts restrictions on the physically reasonable range of torsion and EoS parameters for which trapped surfaces are failed to form.  In this sense, there exists a lower bound on the $s$ parameter for each EoS so that trapped surface formation does not take place during the collapse scenario. The behavior of the ratio $2{\mathcal E}/R\phi\mathcal{P}$ in terms of the proper time $\tau$ is shown in the upper panels of figure (\ref{Fig2}). Solutions for which the spacetime is un-trapped through the whole duration of
the collapse occur for some negatively pressured ordinary matter contents
for admissible values of the torsion parameter. The upper left panel is plotted for these values of $s$ and $w$ parameters chosen from the shaded region of figure (\ref{regionn}), for which, trapped surfaces are avoided till the singularity formation. As the upper right one shows the apparent horizon forms at the time $\tau_{{\rm AH}}$. Whether the apparent horizon forms sooner or later depends on the dynamics of the collapse scenario.  To clarify the situation let us have a more detailed look at the rate of change of the speed of collapse. From (\ref{H}), the rate of the collapse can be easily read off as
\bea
H=\f{1}{\delta}\f{\dot{\phi}}{\phi},
\eea
whence one has
\be
\ddot{a}=\f{d}{dt}(aH)=\f{1}{\delta}\left[\dot{a}\f{\dot{\phi}}{\phi}+a\left(\f{\ddot{\phi}}{\phi}-\f{\dot{\phi}^2}{\phi^2}\right)\right].
\ee
The above equation can be re-written as
\be\label{rate}
\ddot{a}-\f{\dot{a}}{\delta}\f{\dot{\phi}}{\phi}-\f{1}{\delta}\left(\f{\ddot{\phi}}{\phi}-\f{\dot{\phi}^2}{\phi^2}\right)a=0.
\ee
The coefficient of $\dot{a}$ plays the role of a frictional or anti-frictional term depending on its sign to be positive or negative, respectively. 
For $\delta>0$, the  time derivative of the torsion potential is negative, $\dot{\phi}<0$. Therefore this term acts as friction, making the collapse 
to decelerate and delays the formation of the apparent horizon. 
The lower left panel in figure (\ref{Fig2}) shows the behavior of the apparent horizon time as a function of  $s$ for different values of the EoS parameter. It is seen that the larger the value of the torsion parameter, the longer the time of formation of apparent horizon. On the other hand, for $w<0$, the more negative the pressure of perfect fluid matter, the later the apparent horizon forms (family of black curves). Conversely, for $w>0$, the more the positive pressure the sooner the apparent horizon forms to cover the singularity (family of red curves). Furthermore, as the torsion parameter goes to infinity, all the curves settle on constant lines, i.e., the GR limit. It should be noticed that in a homogeneous collapse setting all the shells of matter distribution become singular at the time at which the physical radius of the collapsing object vanishes, i.e., $R(\tau_{s},r)=0$. However, the behavior of the apparent horizon as is controlled by the equality in (\ref{ratio-ps}), is quite different in the sense that, it can intersect with a typical shell and moves outward to finally disappear or conversely it can move inward to cover the resulted singularity \cite{MALAF}. The location of the apparent horizon is given by
\be\label{EQUL}
2{\mathcal E}(\tau,r_{{\rm Ah}}(\tau))=\phi(\tau){\mathcal P}R(\tau,r_{{\rm AH}}(\tau)),
\ee
which gives the apparent horizon curve as
\be\label{APHC}
r_{{\rm AH}}(\tau)=\f{1}{\alpha}\left[a_i^{-\gamma}+\alpha\gamma(\tau_i-\tau)\right]^{-\f{\delta+3w+1}{2\gamma}}.
\ee
The behavior of apparent horizon curve is sketched in the lower right panel of figure (\ref{Fig2}). The family of blue curves (plus the red one) have been plotted for those values of torsion and EoS parameters chosen from the shaded region of figure (\ref{regionn}). For these values the apparent horizon diverges while as the family of black curves (the values $s$ and $w$ selected from the gray region of figure (\ref{regionn})) show, the apparent horizon converges to finally cover the singularity. In figure (\ref{WECF}) we have plotted the time behavior of the effective pressure, the effective energy density and the second inequality of (\ref{WEC}). As is seen in the upper left panel, except for those values of $s$ and $w$ for which trapped surfaces can be avoided (the shaded region in figure (\ref{regionn})), the effective pressure is positive. The effective energy density remains positive and diverges at the singularity (see the right panel) and finally the second inequality of (\ref{WEC}) always holds (lower left panel). The possibility of violation of the weak energy condition is observed in the collapse settings where the effects of quantum gravity are treated \cite{MALAF,LQGSINA} and recently in noncommutative
collapse settings \cite{MHJP}. In the late stages of the collapse where the energy density becomes many
orders of magnitude larger than the earlier stages of the collapse (comparable to Planck energy) and the length scale is much smaller than the earlier radii of a typical star, the effects of quantum gravity
induces negative pressure that leads to a bounce and thus singularity avoidance, see e.g. \cite{QGSA}.
However, as extensively discussed in \cite{RoBe}, in the classical limit, it is not physically plausible to remove the spacetime singularity at the price of violating the weak energy condition.

So far, we have found a class of collapse solutions where the apparent horizon is avoided till the singularity formation. However, for the homogeneous collapse setting as studied here, the absence of apparent horizon is a necessary but not a sufficient condition for the nakedness of the singularity. As discussed in \cite{Wald-Iyer} there can exist regular Cauchy surfaces which come arbitrary close to the spacetime singularity so that trapped surfaces are avoided prior to these surfaces, though the singularity is not naked. Such a situation has also been reported in \cite{JSITR} where it is shown that in the gravitational collapse of an inhomogeneous dust cloud, the absence of apparent horizon does not necessarily mean that the singularity is indeed naked.
This is in contradiction to the arguments stated in \cite{SHTE}, where numerical simulations suggest the criterion for nakedness of the singularity as the failure of apparent horizon formation on a sequence of spacelike surfaces developing throughout the spacetime. In order to examine the nature of the singularity we must seek whether there can exist any null geodesic terminating in the past at the singularity and reaching  an observer at the boundary of the collapsing cloud. These family of geodesics have to satisfy $d\tau/dr = a(\tau)$ in the interior region so that the area radius must increase along these trajectories. As discussed in \cite{GI}, such a situation cannot take place since all the collapsing shells become singular simultaneously. In other words, the condition on radial null geodesics implies $d\tau/dr=0$ at the singularity time due to the homogeneity. However, this situation could happen after a suitable matching to an exterior region, whose boundary $r = r_{\rm b}$ is the surface of the collapsing cloud that becomes singular at $\tau = \tau_s$, into which null geodesics can escape. We shall treat this issue in the next section where we show that, from the exterior perspective, upon using a standard procedure as developed in \cite{Joshi}, there exists a family of outgoing null rays reaching asymptotic observers to whom the singularity is exposed.

Finally we compute the curvature strength of the naked singularity which is an important aspect of its physical seriousness. When a collapsing star develops a curvature singularity, the energy
density diverges. However, finite physical volumes may or may not be crushed
to zero volume as the singularity is reached. This could be used as a criterion for judging
the physical seriousness of the singularity, and also for the possible extendibility of spacetime
through the singularity. The singularity is said to be gravitationally strong in the sense of Tipler \cite{TIP} if every collapsing volume element is crushed to zero size at the singularity, otherwise it is known as weak.
It is believed that spacetime cannot be extended through a strong singularity, but is possibly extendable through a weak one \cite{CL}. Clarke and Krolak \cite{CLKR} gave necessary and sufficient quantitative
criteria for the singularity to be gravitationally strong, by examining the rate of curvature growth along outgoing non-spacelike geodesics terminating at the singularity. In order that the singularity be gravitationally strong there must exist at least one non-spacelike geodesic with tangent vector $\ell^{\mu}$, along which the following condition holds in the limit of approach to the singularity
\be\label{CSRC}
\Upsilon=\lim_{\substack{k\rightarrow 0}} k^2R_{\mu\nu}\ell^{\mu}\ell^{\nu}>0,
\ee
where $R_{\mu\nu}$ is the Ricci tensor and $k$ is an affine parameter which vanishes at the singularity. A geodesic is an extremal curve $x^{\mu}=x^{\mu}(k)$ satisfying
\be\label{EXTCU}
\delta\int ds=\delta\int\left[\psi^2g_{\mu\nu}\f{dx^{\mu}}{dk}\f{dx^{\nu}}{dk}\right]^\f{1}{2}dk=0.
\ee
The Euler-Lagrange equations for the geodesics turn out to be \cite{Soleng}
\be\label{ELGE}
\psi\f{d\ell^{\mu}}{dk}+\psi\{^\mu\,_{\alpha\beta}\}\ell^{\alpha}\ell^{\beta}-\ell^{\alpha}\ell_{\alpha}\psi^{,\mu}+2\ell^{\mu}\ell^{\alpha}\psi_{,\alpha}=0.
\ee
Let us consider radial null geodesics with the tangent vector $\ell^{\mu}=dx^{\mu}/dk=\left[\ell^{\tau},\ell^r,0,0\right]$ that terminate at the singularity at $k=0$. These geodesics have to satisfy
\bea\label{NGE}
\f{d\ell^{\tau}}{dk}+\{^\tau\,_{\alpha\beta}\}\ell^{\alpha}\ell^{\beta}+2\left(\ell^{\tau}\right)^2\f{\dot{\psi}}{\psi}&=&0,\nn
\f{d\ell^{r}}{dk}+\{^r\,_{\alpha\beta}\}\ell^{\alpha}\ell^{\beta}+2\ell^{\tau}\ell^{r}\f{\dot{\psi}}{\psi}&=&0,
\eea
where we have used the null condition $\ell^{\mu}\ell_{\mu}=0$. Substituting then for the solution and non-vanishing components of Christoffel symbols the resulting differential equations can be easily integrated to give
\be\label{ztzr}
\ell^{\mu}=\left[a^{-(\delta+1)},a^{-(\delta+2)},0,0\right].
\ee
In order to calculate the null curvature condition (NCC) $R_{\mu\nu}\ell^{\mu}\ell^{\nu}> 0$, we project the Ricci tensor along the null vector field given above to get
\bea\label{PROJ}
R_{\mu\nu}\ell^{\mu}\ell^{\nu}&=&\f{8\pi\rho_{i_{m}}}{\phi}(1+w)a^{-(3(1+w)+2(1+\delta))}+\left[\f{3}{2}(s^2-1)\f{\dot{\phi}^2}{\phi^2}-2H\f{\dot{\phi}}{\phi}+2\f{\ddot{\phi}}{\phi}\right]a^{-2(1+\delta)},\nn
&=&\frac{\alpha^2}{2}\delta \left[2(\gamma-1) +\delta(3 s^2-1)\right]a^{2 (\gamma -\delta -1)}+\frac{8\pi\rho_{i_{m}}(w+1)}{\beta } a^{-(5+3\delta+3w)},
\eea
where use has been made of  equations  (\ref{Einstein}) and (\ref{H}). The lower right panel of figure (\ref{WECF}) shows the behavior of NCC as a function of scale factor where it is seen that this condition is satisfied throughout the collapse scenario. The fulfillment of NCC is crucial in the satisfaction of {\it limiting strong focusing condition} (LSFC) which is regarded as a criterion for its destructiveness property of any object that falls into it\footnote{It has also been pointed out in \cite{QPCNSC} that the curvature strength of a naked singularity is related to its quantum effects so that in the case of strong naked singularity formation, quantum particle creation diverges in the limit of approaching the Cauchy horizon.} \cite{CLKR}. Next in order to compute the LSFC we proceed by noting that
\be\label{ddk}
\f{d}{dk}a^{\epsilon}=\epsilon Ha^{\epsilon-1},~~~~~~\f{d^2}{dk^2}a^{\epsilon}=\epsilon\left[(\epsilon-1)H^2+aHH_{,a}\right]a^{\epsilon-2},
\ee
whereby using the last part of (\ref{PROJ}) we finally get
\bea\label{UPFFF}
\Upsilon&=&\lim_{\substack{k\rightarrow 0}} k^2R_{\mu\nu}\ell^{\mu}\ell^{\nu}=\lim_{\substack{k\rightarrow 0}}k^2\left[\f{\alpha^2\delta \left[2(\gamma-1) +\delta(3 s^2-1)\right]}{2a^{2(\delta-\gamma+1)}}+\f{8\pi\rho_{i_{m}}(w+1)}{\beta a^{3(\delta+w)+5}}\right]\nn
&=&\Delta_1a^{-2\delta}+\Delta_2a^{-\left[2\gamma +3( \delta + w+1)\right]},
\eea
where
\be\label{COFFS}
\Delta_1=\frac{\delta\left(2\gamma+\delta(3s^2-1)-2\right)}{2(\gamma -2\delta -1) (\gamma -\delta -1)},~~~~~~~\Delta_2=\frac{16 \pi\rho_{i_{m}}(w+1)}{\alpha^2 \beta \left(3(\delta+w)+5\right) (4 + 3(\delta+w) +\gamma)}.
\ee
In order that the condition $\Upsilon>0$ be satisfied, we require $\Delta_1>0$ and $\Delta_2>0$. The fulfillment of the new conditions on the curvature strength of the naked singularity makes the shaded region in figure (\ref{regionn}) to be restricted as the left panel of figure (\ref{SPR}) shows. The shaded region is the original one for which the trapped surfaces are avoided requiring the physical reasonableness. The dotted region shows the region where the LSFC is satisfied. The region where $\Upsilon<0$ is unphysical since it violates the weak energy condition. Furthermore, we note that the initial density profile has to be non-negative as WEC requires, and thus the positivity of the coefficients $\Delta_1$ and $\Delta_2$ depend only on the EoS and the torsion parameter. Therefore, we conclude that the naked singularity is Tipler strong independently of the initial profile of energy density.


\section{Exterior solution}\label{ExtS}
The gravitational collapse setting studied so far deals with the interior of the collapsing object.
In order to complete the model we need to match the interior spacetime to a suitable exterior one.
From a conceptual viewpoint the surface of a collapsing star divides the whole spacetime into two regions; 
the interior region, filled with matter and radiation and the exterior region which is usually filled  with 
all types of radiation coming out of the star. Since, on one hand, the exterior region of a realistic astrophysical 
object is surrounded by a radiation zone and on the other, the internal radial pressure is nonzero at the boundary, 
the Schwarzschild metric may no longer be a good approximation to describe the external region of such a star. 
It should be modeled by more suitable spacetimes, namely the Vaidya spacetime \cite{Vaidya}. The Vaidya solution 
has been vastly studied in gravitational collapse \cite{JOGLO} and particularly to describe the formation of naked 
singularities \cite{VAINS}. In the presence of a cosmological constant, it has been shown that the collapse of the 
spherically symmetric matter bulk will result in naked singularities rather than black holes \cite{LEM}.

\begin{figure}
\includegraphics[scale=0.5]{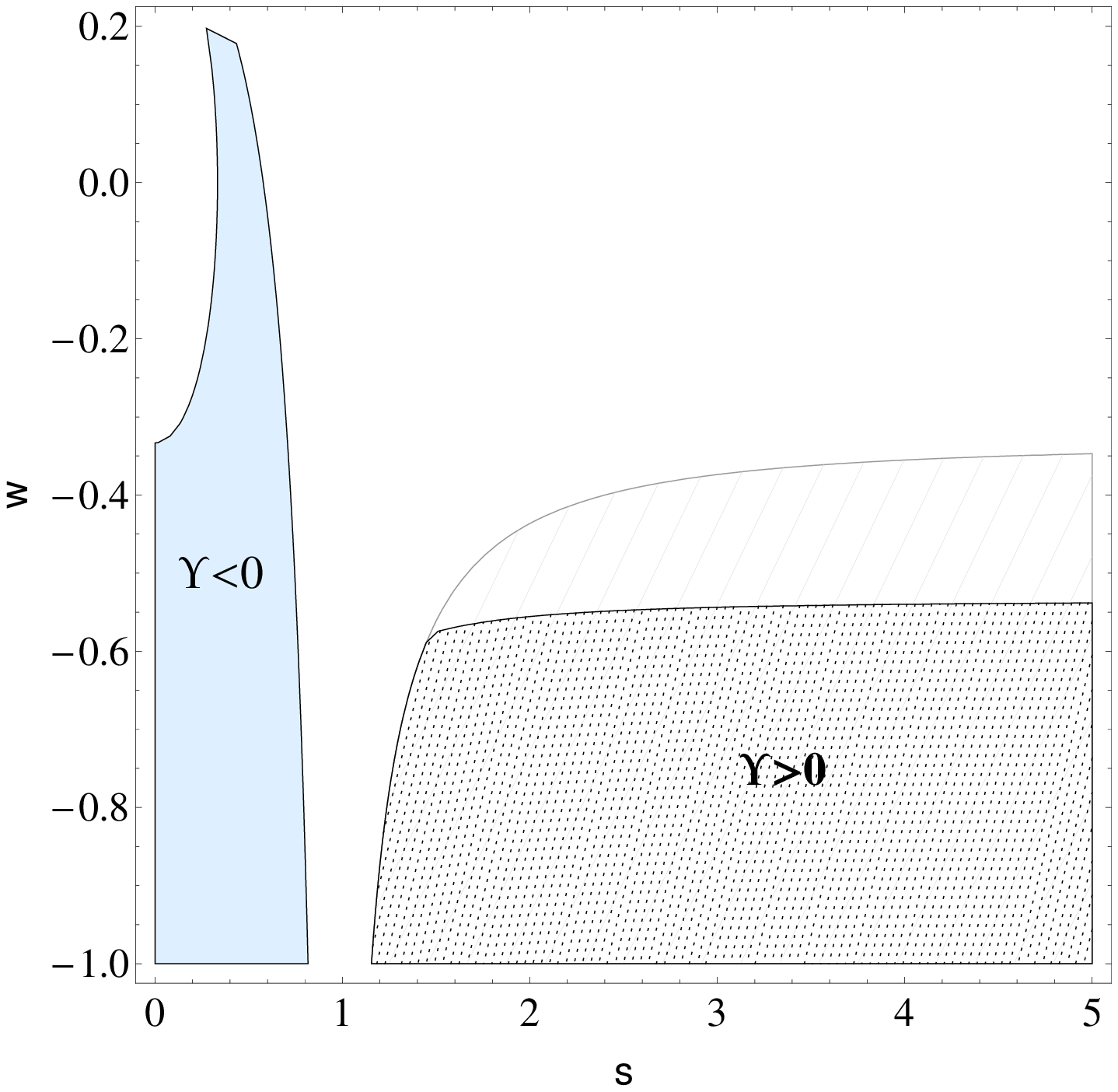}\hspace{.51cm}
\includegraphics[scale=0.5]{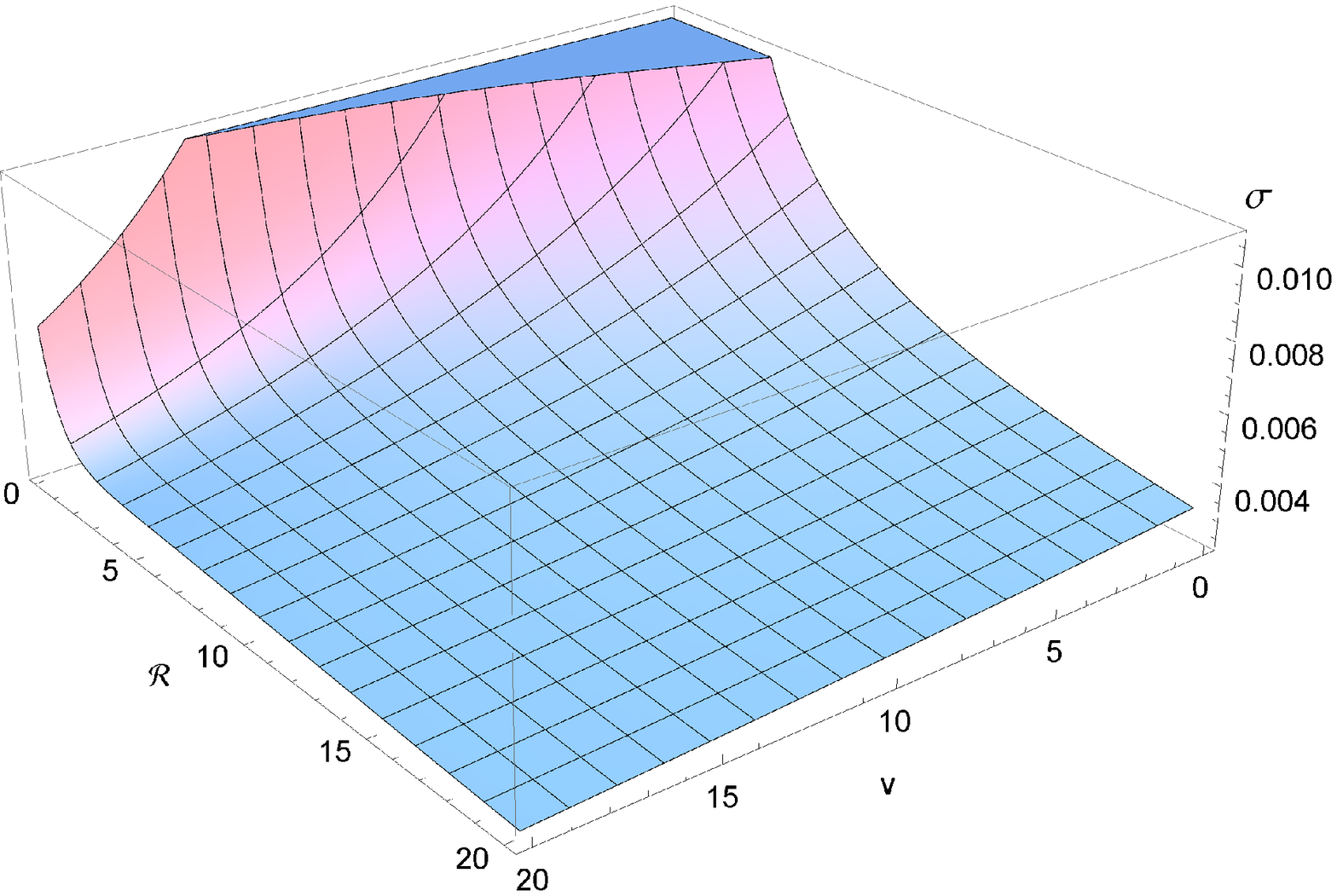}
\caption{Left panel: The allowed region for torsion strength and EoS parameters so that the naked singularity be gravitationally strong (dotted region). The light blue region is not allowed since it violates the weak energy condition. Right panel: The behavior of energy versus retarded null coordinate and the distance for $s=1.24$, $w=-0.8$ and $C_0=10$. }\label{SPR}
\end{figure}

Let us take the exterior spacetime line element as that of generalized Vaidya metric \cite{GVMEX} which in retarded (exploding) null coordinates is given by
\begin{equation}\label{extmet}
ds^2_{+}=-f({\mathcal R},v)dv^2-2dvd{\mathcal R}+{\mathcal R}^2d\Omega^2,
\end{equation}
where $f({\mathcal R},v)=1-2{\rm M}(\mathcal R, v)/{\mathcal R}$ is the exterior metric function, with ${\rm M}(\mathcal R, v)$ being a measure of the mass contained within the radius ${\mathcal R}$. We label the radiation coordinates as $\{X_+^{\mu}\}\equiv\left\{v,{\mathcal R},\theta,\phi\right\}$ with $v$ being the retarded null coordinate labeling different shells of radiation and ${\mathcal R}$ being the Vaidya radius. Utilizing the Israel-Darmois junction conditions \cite{IDJCC}, we are to match the interior line element
\begin{equation}\label{intel}
ds_{-}^2=-d\tau^2+a^2(\tau)dr^2+a^2(\tau)(r^2d\theta^2+r^2\sin^2\theta d\phi^2),
\end{equation}
to the exterior metric given above through the boundary surface $r=r_{\rm b}$. The interior coordinates are labeled as $\{X_-^{\mu}\}\equiv\{{\tau,r,\theta,\phi}\}$. The interior and exterior induced metrics take the form, respectively
\be\label{ININ}
ds_{{\rm b}^{-}}^2=-d\tau^2+a^2(\tau)r_{b}^2(d\theta^2+\sin^2\theta d\phi^2),
\ee
and
\be \label{INEX}
ds_{{\rm b}^{+}}^2=-\left[f\big({\mathcal R}(\tau),v(\tau)\big)\dot{v}^2+2\dot{{\mathcal R}}\dot{v}\right]d\tau^2-{\mathcal R}^2(\tau)(d\theta^2+\sin^2\theta d\phi^2).
\ee
Matching the induced metrics gives
\be\label{FFF1}
f\big({\mathcal R}(\tau),v(\tau)\big)\dot{v}^2+2\dot{{\mathcal R}}\dot{v}=1,~~~~~~{\mathcal R}(\tau)=r_{\rm b}a(\tau),
\ee
where an overdot denotes $d/d\tau$. In order to find the extrinsic curvatures of the interior and exterior hypersurfaces close to the boundary surface we need to find the unit normal  vector fields to these hypersurfaces. A straightforward calculation reveals that
\be\label{NVF+-}
n^{-}_{\mu}=\left[0,a(\tau),0,0\right],~~~~~~n^{+}_{\mu}=\frac{1}{\left[f({\mathcal R},v)\dot{v}^2+2\dot{{\mathcal R}}\dot{v}\right]^{\frac{1}{2}}}\left[-\dot{\mathcal R},\dot{v},0,0\right].
\ee
The components of extrinsic curvature for interior region read
\be\label{EXCIN}
K_{\tau\tau}^-=0,~~~~~K_{\theta}^{\theta-}=K_{\phi}^{\phi-}=\f{1}{{\mathcal R}\psi},~~~K_{\tau\theta}^-=K_{\tau\phi}^-=K_{\theta\phi}^-=0.
\ee
For the exterior region we have
\bea\label{EXCEX}
K_{tt}^{+}&=&-\frac{\dot{v}^2\left[ff_{,{\mathcal R}}\dot{v}+f_{,v}\dot{v}+3f_{,{\mathcal R}}\dot{{\mathcal R}}\right]+2\left(\dot{v}\ddot{{\mathcal R}}-\dot{{\mathcal R}}\ddot{v}\right)}{2\left(f\dot{v}^2+2\dot{{\mathcal R}\dot{v}}\right)^{\frac{3}{2}}},\nonumber\\
K^{+\theta}_{\theta}&=&K^{+\phi}_{\phi}=\frac{f\dot{v}+\dot{{\mathcal R}}}{{\mathcal R}\sqrt{f\dot{v}^2+2\dot{{\mathcal R}}\dot{v}}}.
\eea
We assume that there is no surface stress-energy associated with the boundary surface (see e.g. \cite{SURLAYER} for the study of junction conditions for boundary surfaces and surface layers). For a smooth matching of extrinsic curvatures in the absence of surface layer, we find
\bea
f\dot{v}+\dot{{\mathcal R}}&=&\psi^{-1},\label{TETE}\\
\dot{v}^2\left[(ff_{,{\mathcal R}}+f_{,v})\dot{v}+3f_{,{\mathcal R}}\dot{{\mathcal R}}\right]&+&2\left(\dot{v}\ddot{{\mathcal R}}-\dot{{\mathcal R}}\ddot{v}\right)=0.\label{TT}
\eea
Taking derivatives of (\ref{TETE}) and the first part of (\ref{FFF1}), we can construct the following expressions
\bea\label{EXPd}
2\dot{{\mathcal R}}\ddot{v}&=&\dot{f}\dot{v}^2+2\dot{v}\dot{\psi}\psi^{-2}\nn
\ddot{{\mathcal R}}\dot{v}&=&-\ddot{v}(f\dot{v}+2\dot{{\mathcal R}})+\dot{v}\dot{\psi}\psi^{-2}.
\eea
Substituting the above expressions into the last term of (\ref{TT}) and after a little algebra we get
\be\label{F,V}
f_{,v}=-\f{2\dot{\psi}}{\dot{v}^2\psi^{2}}.
\ee
Finally from the first part of (\ref{FFF1}) and (\ref{TETE}) we get the four velocity of the boundary as
\be\label{4-V}
U^{\mu}=\left(\dot{v},\dot{{\mathcal R}},0,0\right)=\left[\f{1+\sqrt{1-f\psi^2}}{f\psi},-\f{\sqrt{1-f\psi^2}}{\psi},0,0\right],
\ee
where the minus sign for $\dot{{\mathcal R}}$ is chosen since we are concerned with a collapse setting. In order to find the exterior metric function we proceed by noting that
\be\label{psidotpsi}
\dot{\psi}=\f{\sqrt{\beta}\delta}{2}r_{\rm b}^{-\f{\delta}{2}}{\mathcal R}^{\f{\delta}{2}-1}\dot{{\mathcal R}},~~~~\psi^2=\beta\left(\f{{\mathcal R}}{r_{\rm b}}\right)^{\delta},
\ee
whence, using (\ref{4-V}) we arrive at the following partial differential equation for the exterior metric function
\be\label{PADIFF}
\f{\partial}{\partial v}f({\mathcal R},v)=\f{\delta}{{\mathcal R}}\left\{\f{f({\mathcal R},v)^2\left[1-\beta f({\mathcal R},v)\left(\f{{\mathcal R}}{r_{{\rm b}}}\right)^{\delta}\right]^{\f{1}{2}}}{\left[1+\sqrt{1-\beta f({\mathcal R},v)\left(\f{{\mathcal R}}{r_{{\rm b}}}\right)^{\delta}}\right]^2}\right\},
\ee
for which the solution reads
\be\label{solex}
f({\mathcal R},v)=-\f{4\left[\delta {\mathcal R} v+{\mathcal R}^2\left(\beta\left(\f{{\mathcal R}}{r_{{\rm b}}}\right)^{\delta}+g({\mathcal R})\right)\right]}{\left[\delta v+{\mathcal R}g({\mathcal R})\right]^2},
\ee
with $g({\mathcal R})$ being an arbitrary function. The existence of naked singularity can be worked out by considering the behavior of null rays near the singularity. The equations for radial null geodesics emanating from the singularity are given by
\bea\label{RNO}
\f{d}{d\lambda}\xi^v&-&\f{1}{2}f_{,{\mathcal R}}(\xi^v)^2=0,\nonumber\\
\f{d}{d\lambda}\xi^{{\mathcal R}}&+&\f{1}{2}\left(f_{,v}+ff_{,{\mathcal R}}\right)
(\xi^{v})^2+f_{,{\mathcal R}}\xi^{v}\xi^{{\mathcal R}}=0,
\eea
with $\xi^{\mu}=\left[\xi^v,\xi^{{\mathcal R}},0,0,\right]=\left[dv/d\lambda,d{\mathcal R}/d\lambda,0,0\right]$ being the tangent vector of the null trajectory and $\lambda$ being the affine parameter. An outgoing radial null geodesic that meets the singularity in the past has to satisfy the null condition $\xi^{\mu}\xi_{\mu}=0$ or equivalently $ds^2_{+}=0$, i.e.,
\be\label{NULLC}
\f{dv}{d{\mathcal R}}=-\f{2}{f({\mathcal R},v)}.
\ee
Let us define $Z=v/{\mathcal R}$ to be the tangent to the outgoing radial null geodesics from the singularity. If such a tangent to the null geodesics do exist in the limit of approach to the singularity, i.e., ${\mathcal R}\rightarrow 0$ and $v\rightarrow 0$
we must have\footnote{We note that the Kretschmann scalar computed by the metric (\ref{extmet})
\be\label{KRSC}
K=\f{1}{{\mathcal R}^4}\left[\left(\f{\partial^2f({\mathcal R},v)}{\partial {\mathcal R}^2}\right)^2R^4+4\left(\f{\partial f({\mathcal R},v)}{\partial {\mathcal R}}\right)^2{\mathcal R}^2+4-8f({\mathcal R},v)+4f({\mathcal R},v)^2\right],
\ee
diverges at the singularity point $({\mathcal R},v)=(0,0)$.}
\be\label{Z0}
Z_0=\lim_{\substack{{\mathcal R}\rightarrow 0\\ v\rightarrow 0}}\f{v}{{\mathcal R}}=\lim_{\substack{{\mathcal R}\rightarrow 0\\ v\rightarrow 0}}\f{dv}{d{\mathcal R}}=\lim_{\substack{{\mathcal R}\rightarrow 0\\ v\rightarrow 0}}\f{-2}{f({\mathcal R},v)}.
\ee
The central shell-focusing singularity is naked if the above equations admit one or more positive real roots $Z_0$; in this case, there is at least one radially outgoing null geodesic originating at the singularity \cite{JOGLO}. For the exterior metric function obtained in (\ref{solex}), with the free function taken as $g({\mathcal R})=C_0$, we have
\be\label{Z00}
Z_0=\lim_{\substack{{\mathcal R}\rightarrow 0\\ v\rightarrow 0}}\f{\left[C_0 +\delta Z\right]^2}{2\left[\delta Z+\beta\left(\f{{\mathcal R}}{r_{{\rm b}}}\right)^{\delta}+C_0\right]},
\ee
where we note that $\delta>0$ in order to have finite values for $Z_0$ \cite{HarkoST}. Taking the limit we obtain the following algebraic equation that governs the behavior of the tangent to radial null geodesics near the singularity
\be\label{RZ0}
\delta(2-\delta)Z_0^2+2C_0(1-\delta)Z_0-C_0^2=0,
\ee
for which we readily get the solutions as
\bea\label{QASOL}
Z_0^+=-\f{C_0}{\delta},~~~~~~~Z_0^-=-\f{C_0}{\delta-2}.
\eea
Therefor, respecting the condition $\delta>0$, if we take $C_0<0$ there can be found outgoing radial null geodesics emerging from the singularity, with positive definite tangent and exposing it to faraway observers. Such a family of outgoing radial null geodesics can also be found for $C_0>0$ and $0<\delta<2$. The flux of outgoing radiation is given by \cite{FAYFL}
\be\label{EDNR}
\sigma=-\f{2}{{\mathcal R}^2}\f{\partial {\rm M}}{\partial v}=\f{4\delta\left[2\beta{\mathcal R}\left(\f{{\mathcal R}}{r_{{\rm b}}}\right)^{\delta}+C_0{\mathcal R}+\delta v\right]}{(\delta v+C_0 \mathcal{R})^3}.
\ee
Since the star is radiating energy to the exterior zone due to the appearance of negative pressure in the interior, we must have, $\dot{{\rm M}}=\f{\partial {\rm M}}{\partial v}<0$. The right panel in figure (\ref{SPR}) shows the outflow of radiation as a function of retarded time and distance. The flux decreases monotonically as both the null coordinate and radius increase and vanishes asymptotically. The solution (\ref{solex}) reduces to a static spacetime in the GR limit where $\delta\rightarrow0$. In this situation, though the exterior spacetime is static, null geodesics may still have a chance to escape from the singularity, in the sense that if the collapse velocity is bounded the boundary surface $r=r_{{\rm b}}$ can be taken sufficiently small so that no horizon can develop to cover the singularity \cite{GIAMKH}.



\begin{figure}
\includegraphics[scale=0.6]{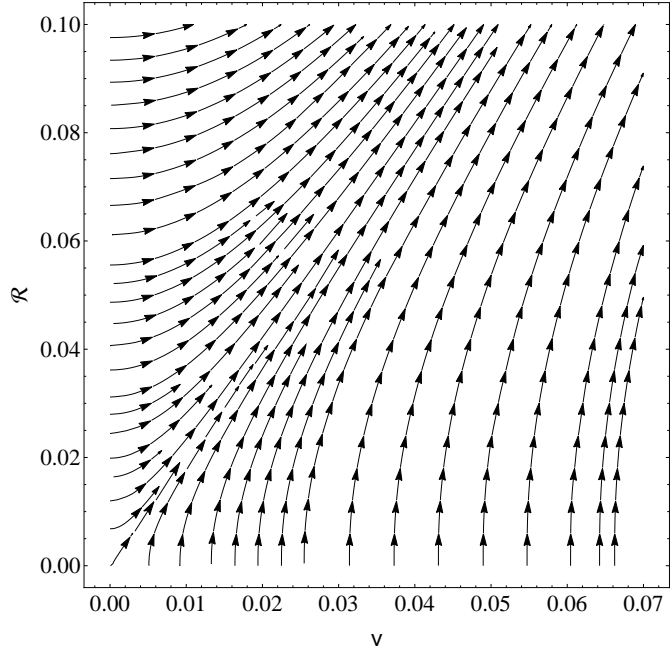}
\caption{Phase portrait in the $({\mathcal R},v)$ plane for $B=-0.1$ and $g(0)=-0.1$.}\label{SPR1}
\end{figure}

At the close of this section, we would like to mention that in the above study, we have taken the free function to be a constant value. To investigate the nature of the singularity for the general form of the free function, we need to examine the behavior of null rays in the vicinity of the point ${\mathcal R}=0$, $v=0$. Let us now assume that this function is well-defined near the singular point and expand the exterior metric function as a Taylor series \big(we take $\delta=1$\big)
\bea\label{EXEX}
f(R,v)\bigg|_{\substack{{\mathcal R}\rightarrow 0\\ v\rightarrow 0}}=-\f{4{\mathcal R}}{v}&+&\f{4g(0){\mathcal R}^2}{v^2}+\left(4\f{g^{\prime}(0)-\beta}{v^2}-4\f{g(0)^2}{2v^3}\right){\mathcal R}^3+4g(0)^3\f{{\mathcal R}^4}{v^4}+8g(0)\left(\beta-g^{\prime}(0)\right)\f{{\mathcal R}^4}{v^3}\nn
&+&2\f{g^{\prime\prime}(0)}{v^2}{\mathcal R}^4+{\mathcal O}(R^5)+{\mathcal O}(v^5),
\eea
where the prime denotes total derivative with respect to ${\mathcal R}$. Taking the first derivative of the free function to be $g^{\prime}(0)=\beta$ and its second derivative to be zero we find that the spacetime behaves self-similarly in the vicinity of the singularity and admits a homothetic Killing vector field given by \cite{JD}
\be\label{HKVF}
X^{\mu}={\mathcal R}\f{\partial}{\partial {\mathcal R}}+v\f{\partial}{\partial v},
\ee
which satisfies the condition $\pounds_{\!X}g_{\mu\nu}=2g_{\mu\nu}$. Therefore, a conserved quantity along the radial null geodesics can be found as
\be\label{CQU}
X^{\mu}\xi_{\mu}={\mathcal R}\xi_{{\mathcal R}}+v\xi_v=constant.
\ee
Following \cite{JD}, we take $\xi^{v}=Q({\mathcal R},v)/{\mathcal R}$, whereby  we can find the first integral of geodesic equations (\ref{RNO})
\be\label{RNORE}
\xi^{v}=\f{dv}{d\lambda}=\f{Q({\mathcal R},v)}{{\mathcal R}},~~~\xi^{{\mathcal R}}=\f{d{\mathcal R}}{d\lambda}=-\f{f({\mathcal R},v)}{2{\mathcal R}}Q({\mathcal R},v),
\ee
where we have used the null condition and $Q$ satisfies the following equation
\be\label{QEQ}
\f{dQ}{d\lambda}+\f{Q^2}{2{\mathcal R}^2}\left(f-{\mathcal R}f_{,{\mathcal R}}\right)=0.
\ee
From the condition (\ref{CQU}) we arrive at an algebraic equation for $Q$, whence we finally get
\bea\label{QSOL}
\f{dv}{d\lambda}&=&B\left[\f{2g(0){\mathcal R}^2}{v}\left(1-g(0)\f{{\mathcal R}}{v}+g(0)^2\f{{\mathcal R}^2}{v^2}\right)-{\mathcal R}\right]^{-1},\\
\f{d{\mathcal R}}{d\lambda}&=&-2B\left[-\f{{\mathcal R}}{v}+g(0)\f{{\mathcal R}^2}{v^2}-g(0)^2\f{{\mathcal R}^3}{v^3}+g(0)^3\f{{\mathcal R}^4}{v^4}\right]\left[\f{2g(0){\mathcal R}^2}{v}\left(1-g(0)\f{{\mathcal R}}{v}+g(0)^2\f{{\mathcal R}^2}{v^2}\right)-{\mathcal R}\right]^{-1},
\eea
where $B$ is a constant. Figure (\ref{SPR1}) shows the numerical plot of the vector fields \big($\xi^{v},\xi^{{\mathcal R}}$\big) in the $({\mathcal R}-v)$ plane, for fixed values $g(0)$ and $B$. We see that the singularity ${\mathcal R}=0$, $v=0$ is a source and there exists at least one radial null geodesic emerging from the singularity with finite slope and escaping to future null infinity ($\scri^+$). The first null trajectory $v=0$ leaves the singularity to reach $\scri^+$ and the emergence of the rest with $v>0$ is followed afterwards. The singularity is also globally naked since $d{\mathcal R}/dv>0$ as $v$ increases. It should be noted that although the two vector fields diverge at $({\mathcal R},v)=(0,0)$, the tangent to null trajectories is positive and finite.
\section{Concluding remarks}
The process of gravitational collapse of a massive body and its final fate have received considerable attention in recent years due to its importance in black hole physics and the issue of cosmic censorship conjecture \cite{CCC}. 
Such studies have provided modern examples of naked singularities as the possible end state of a collapse setting under a variety of circumstances. Recent developments in gravitational collapse physics suggest that naked singularities arising from the exact solutions to Einstein's field equations can take different configurations. Not surprisingly, the observational aspects of such objects should be diverse and interesting. If these objects do exist in nature, it would be decisive in any astrophysical attempt to investigate how they may interact with their surroundings in order to understand whether they can be observed and how (see e.g. \cite{RECJO} and references therein). As discussed in \cite{NSVSBH} a naked singularity, if present, can be observationally distinguished from a Schwarzchild black hole because of the physical properties of their accretion disks which may form around them. In addition, gravitational lensing effects due to extreme curvature regions provide one with a suitable tool to seek the observational signatures coming out from a naked singularity in such a way that lensing characteristics of such objects are qualitatively very different from those formed as Schwarzchild black holes \cite{VirEllis}. From this viewpoint, It is interesting to consider the equation of photon trajectories  for the exterior metric function given in (\ref{solex}) and investigate the different aspects of the lensing effects of the strong curvature naked singularities in the herein model as compared to those reported in the literature (see \cite{VirEllis} and references therein). However, dealing with this issue is out of the scope of the present work.

In the past decades, a large amount of work has been devoted to develop theories of gravity which,  contrary to Einstein-Cartan theory, allow the long range torsion mediated interactions and include a scalar field as the potential of the spacetime torsion \cite{BDLyra}, \cite{TPAG}. Possible experimental consequences of propagating torsion has been studied in \cite{CAR,EXTOR} and particularly the authors in \cite{CAR} concluded that the torsion decays quickly outside of the matter distribution hence leaving no long-range interaction to be discovered experimentally.

However, as a matter of reality hidden behind the most unknown or less-known phenomena in nature, one may not certainly decide on the irrelevance of 
torsion in gravitational physics. While the interaction of matter fields with torsion is weakened at low energy limits\footnote{Theoretical settings for the 
possibility of detection of torsion fields at low energy limits have been investigated in \cite{LOWTOR}}, the essential manifestations of torsion effects may not be negligible in the realm of particle physics \cite{SHAP} or in very early universe \cite{EARTOR}. From such a perspective, the effects of dynamical spacetime
torsion on the final outcome of a gravitational collapse process and possibly its observational features as stated in \cite{NSVSBH} and \cite{VirEllis}, could 
provide a major motivation to investigate this issue since the late stages of the gravitational collapse of a massive star where we are encountered with extreme dense regions could be comparable with the very early stages of cosmological scenarios. Besides the framework of the models mentioned above, Lyra geometry is 
another such a theory which provides a suitable setting for a propagating spacetime torsion and its possible effects on a typical gravitational collapse scenario. In addition to the cosmological models based on this theory \cite{2Singh} we have obtained a class of solutions in this context which describes the process of
gravitational collapse of a massive body. The action integral in Lyra manifold constructed by Lyra curvature scalar can be suitably rewritten as that of 
Brans-Dicke theory  where the BD scalar field arises as a pure geometrical object, a potential for the spacetime torsion. Having taken a spherically symmetric 
homogeneous perfect fluid, obeying a linear equation of state, as the matter content which evolves through the spacetime with a non-zero torsion, we have shown 
that under the satisfaction of certain conditions on physical reasonableness of the collapse, the formation or otherwise of the trapped surfaces is decided by the torsion parameter and EoS as figure (\ref{regionn}) shows. For those class of solutions leading to trapped surface formation, variation of the torsion potential may
behave as a frictional term in the equation governing the rate of collapse (\ref{rate}) so that the collapse continually decelerates; in this sense, the dynamics of the  apparent horizon may be influenced as a consequence of propagating spacetime torsion (see the lower left panel of figure (\ref{Fig2})). On the other hand,
for the second class of solutions where trapped surfaces do fail to form, the pressure turns to be negative in the interior of the collapsing cloud, thus generating an outward flux of energy to the exterior region. One then may intuitively imagine that the collapsing object may loose away some of its matter content
preventing thus the trapped surface formation, since there remains not enough mass at each stage of the collapse to get the light trapped. Furthermore, in 
view of figure (\ref{regionn}), the condition on the EoS for nakedness or otherwise of the singularity in the GR limit is $w<-1/3$ or $w>-1/3$, respectively 
which is in agreement with the result found in \cite{COOP}. Whereas, when we deviate from GR by making the torsion parameter to be sufficiently low\footnote{We note it should not be misunderstood that the large values of the torsion parameter means the more the contribution of the spacetime torsion to the collapse setting. Just as the BD theory, the larger we take the BD coupling parameter the weaker the BD scalar field does play its role in the gravitational interaction.}, trapped 
surfaces can be either formed or fail to form (for $w<-1/3$) depending on the value of the $s$ parameter. This puts  a lower bound on the torsion parameter 
for each value of the EoS as dictated by the red boundary curve.\footnote{Interpolating between the points living on the red boundary curve gives the relation 
between the torsion and EoS parameters as, $w\simeq -0.3333-2.2484{\rm exp}(-1.2747s)$. We note that neither the points in the gray region of figure 
(\ref{regionn}) nor the ones in the shaded region are allowed to lay on this curve.} Despite the fact that the singularity produced as the end state of the 
collapse configuration could be potentially naked for the torsion and EoS parameters as given in the shaded region of figure (\ref{regionn}), it cannot be 
gravitationally strong for all the values of these parameters. The requirement that the LSFC be fulfilled further restricts the possible range of $s$ and $w$ 
parameters which are plotted as the dotted region in the left panel of figure (\ref{SPR}).

The collapse setting studied here deals with the homogeneous density profile while its dynamics develops quite differently once the inhomogeneities are introduced within the configuration of the collapsing object. The simplest and rather specific scenario for the gravitational collapse was that of Oppenheimer-Snyder-Datt (OSD) model which describes the process of gravitational collapse of a homogeneous pressure-less cloud with no rotation and internal stresses \cite{OSD}. It is obvious that employing the OSD model in order to describe the collapse process of a realistic compact object is highly idealized and unphysical. As a matter of fact, we would expect that the density distribution would typically grow when we approach the regions near the center and decreases once we move away from the center. The first step in coming closer to an actual model of collapsing star was to generalize the OSD set up to an inhomogeneous collapsing dust cloud, leading to the celebrated LTB collapse models \cite{LTB}. In this respect, there has been an extensive and long-standing debate on the role of initial data in determining the final outcome of a collapse scenario and more importantly the genericity and stability features for naked singularities and black holes that form as
the end products of a complete gravitational collapse of a massive  cloud of matter (see e.g. \cite{Joshi,RECJO,GSPGS}). In connection with these points, the final state of a collapse process would drastically change in the presence of inhomogeneities, since the horizon behaves differently in contrast to the case where the densities are taken to be completely homogeneous. Therefore, with regard to the model described here, the role of initial profile of the energy density can be only manifested in the satisfaction of the regularity condition which is clearly due to the homogeneous nature of the model. While, in a typical LTB collapse model the initial density distribution is expanded as a Taylor series near the center of symmetry so that it is the derivatives of the initial density profile that play a crucial role in the nakedness and curvature strength of the singularity (see e.g., \cite{Joshi} and references therein). However, the study of the effects of inhomogeneous torsion potential and more interestingly those due to the inclusion of intrinsic angular momentum density of spinning fluids on the collapse dynamics and the time evolution of the apparent horizon could also pave the way for any feasible formulation and proof for the CCC. Such a study is currently under consideration.

There are also a few features that beg some additional elucidation. The relation between torsion and BD coupling parameters can be viewed from two standpoints: in one hand, finding experimental bounds on the torsion parameter would lead to the constraints on the BD parameter. On the other hand, the observational restrictions on the BD coupling parameter can indeed be used to determine the constraints on the torsion parameter. The investigation of renormalizability of quantum field theory in the presence of an external gravitational field with torsion has shown that the interaction of matter fields with torsion has many physical features, so that the renormalization structure of the field theory on the curved spacetime with torsion requires the non-minimal interaction of the external torsion with spinor fields\footnote{The non-minimal interaction of spacetime torsion with fermionic matter has received remarkable attention for the purpose of formulating GR as a gauge theory \cite{GTG}.} and hence the introduction of non-minimal coupling parameters that determine such interaction \cite{REQFT}. Therefore, the methods of the effective field theory has been developed and exploited in order to seek  the theoretical features as well as phenomenological footprints of the spacetime torsion. In this connection, experimental probes on the violation of local Lorentz invariance in the spacetime region where  torsion is non-vanishing is utilized to extract constraints on the components of  torsion \cite{Kostel}. Work along this line has been done through Hughes-Drever experiments \cite{HDE} and in the most recent work, experimental bounds on the components of the torsion field has been reported by the search for parity violation in neutron spin rotation \cite{HE4}.
However, we would like to argue here that the results of \cite{Kostel,HE4} are not applicable to the present model (the bounds on the torsion parameter as given in figure (\ref{regionn})) for two reasons. Firstly, the background torsion field has been taken approximately to be constant while in our model torsion has a dynamical nature. Secondly, the direct coupling of  torsion components to fermion sector in the effective Lagrangian has been taken into account based upon which the constraint on torsion components are found, by reinterpretation of the experiments  seeking for Lorentz and CPT violation; while in the present work the explicit presence of fermion fields and their coupling to the spacetime torsion has not been considered. On the other hand the relation between Lyra torsion parameter and the BD coupling parameter can be notable in the sense that the experimental bounds on the BD parameter would correspondingly confine the torsion parameter. Recent observations of radiation damping in mixed binary systems have put rigorous bounds on the BD coupling parameter as $\omega>40000$ from Cassini measurements of the Shapiro time delay in the Solar-System, $\omega>1000$ from the observational benchmarks of the Nordtvedt effect using lunar laser ranging experiments and  $\omega>1250$, from the observations of the orbital period derivative of the quasi-circular white dwarf neutron star binary \cite{ABWZ}. The authors have estimated the contributions due to tensor and scalar sector of the gravitational radiation from compact binary systems
in the massive BD theory, and used recent observations of radiation damping in mixed binary systems to
put stringent bounds on the parameters of the theory. Post-Newtonian approximation has been exploited to derive the equations of motion in the weak-field limit and then the scalar and tensor gravitational waveforms have been obtained by solving the linearized field equations. In comparison to the constraints on torsion and EoS parameters as specified in figure (\ref{regionn}), we see that the experimental bounds on the BD parameter as reported in \cite{ABWZ} (or within the herein model on the torsion parameter) lead to the result that the related EoS for the apparent horizon formation or its avoidance would deviate negligibly from the GR case. In fact, the Solar-System tests based on Post-Newtonian formalism are reliable in the weak-field limit of gravity, the matter that generates Solar-System gravity moves slowly and has small internal energies. As a consequence, weak-field experiments cannot give much information on strong-field regime, and one needs to test such a regime by resorting to alternative methods. Given that the present values of the Post-Newtonian parameters are obtained from weak-field tests, it is reasonable that deviations from GR can be hidden in the weak-field regime but may become dominant at the early times in the history of the universe or in the regions near the singularity where the spacetime curvatures take extreme values and blow up \cite{PS,CMWill}. It is worth mentioning at this stage that, the BD coupling parameter bears the ratio of the scalar to tensor couplings to matter so that the larger the value of BD parameter the smaller the effects of the scalar field. Therefore the larger value of the BD parameter suggests that the scalar field has a negligible contribution to the gravitational interaction; the theory becomes indistinguishable from GR and correspondingly, from the standpoint of the present model, the effects of spacetime torsion disappear in this limit. However, observational bounds on the present value of the BD coupling parameter do not necessarily put constraints on its value at early times in more general scalar-tensor theories where the BD coupling parameter is allowed to vary \cite{PRD-55-1906}. The generalized scalar-tensor theories in which the $\omega$ parameter depends on the scalar field could have smaller values, therefore significant contribution of the scalar field at the early stages of the universe\footnote{Experimental bounds on the BD coupling parameter obtained by local systems, say from bending of light by the Sun or Solar-System tests may not be necessarily applied  on distances much larger than those of the measurements (those extracted on the cosmological scales) and the epochs much different from the present. Constraints on the BD parameter from the Big-Bang nucleosynthesis demand $\omega\geq32$ \cite{TDBBN} confirmed by the analysis given in \cite{ACBBN}, $\omega>332$ without considering a cosmological constant and $\omega>277$ when including a cosmological constant \cite{TBS} (see also \cite{UZAN} for a recent review).}, while evolving through the attractor mechanism to a large value today \cite{CMWill,VF}. From this viewpoint, whereas the Post-Newtonian formalism provides a uniform description of gravitational interaction in the case of weak fields, there is no such requirement for fields derived on the basis of a Post-Newtonian weak-field approximation in the presence of compact objects with strong gravitational fields (e.g. the final phases of black hole coalescence which involve strong gravitational interactions where only numerical relativity methods are valid or the phenomenon of \lq\lq{}spontaneous scalarization\rq\rq{} for massive neutron stars, see \cite{PS} and references therein) and specially in the collapse setting where super-dense regimes of extreme gravity are dominant. From another viewpoint, the BD theory may be regarded as an approximation for a class of scalar-tensor theories of gravity which have more significant effects on cosmological scales. In this way one can intuitively imagine that the bounds on the torsion parameter (see figure (\ref{regionn})) obtained from those values of $\omega$ parameter at the early universe can be compared to a situation where a dense matter distribution undergoes a collapse process in the presence of a dynamical background torsion so that the effects of a dynamical torsion could alter its final stage.

Finally let us compare our results with \cite{Scheel} where the authors have developed a numerical code that solves the dynamical field equations for evolution of a spherically symmetric gravitational collapse of a pressure-less matter in BD theory. Utilizing this code, they showed that the Oppenheimer–Snyder collapse in this theory results in black hole formation rather than naked singularities, at least for $|3\omega+2|>3$, which are identical to those of GR in final equilibrium, but are quite different during the dynamical evolution.  In the model presented here, motivated by the fact that the BD scalar field can act as a dynamical torsion potential, we studied the collapse of a homogeneous perfect fluid whose pressure and energy density obey a linear EoS. The exact solutions we found show that for physically reasonable values of the EoS and  torsion parameter ($s>1$ compared to the bound on the BD parameter as specified by \cite{Scheel}), both naked singularities and black holes could arise as the collapse end product. Furthermore, the weak and null energy conditions are satisfied in the Jordan representation utilized in this paper which ensure the physical reasonableness of the collapse configuration. The naked singularity formed is gravitationally strong as a consequence of the satisfaction of NCC. Whereas, the NCC is violated in \cite{Scheel}, as a result of which the apparent horizon of the black hole can pass outside of the event horizon with decreasing the surface area
of the event horizon over time. The exterior spacetime of the collapsing object investigated in \cite{Scheel} admits monopole gravitational radiation during its evolutionary stages and settles down to a static Schwarzchild metric at late times where the scalar field approaches a constant value. In our case, the dynamical exterior spacetime is found to be that of the generalized Vaidya spacetime whose dynamical characteristic is due to the mere presence of the gauge function in the Lyra connection being interpreted as the BD scalar field; therefore in Lyra geometry in contrast to Einstein-Cartan theory, the torsion can propagate and thus its effects are transmitted through a smooth matching to exterior region. From the viewpoint of dynamical exterior spacetime, we examined the existence of outgoing radial null geodesics ending in the past at the singularity. It was shown that, there can be found outgoing radial null rays with real positive tangents at the singularity that propagate to faraway observes. This could possibly provide a counterexample to CCC. However, any rigorous formulation
of the CCC depends on a detailed analysis of stability and genericity aspects of the naked singularities and
black hole phases that develop as the end states of a realistic gravitational collapse. In this sense, we do not strictly claim that the CCC is violated here since the resulting naked singularity may not remain stable upon the introduction of perturbations within the matter content and the spacetime geometry, such as the  inhomogeneities that come into play at later stages of the collapse process.


\section*{Acknowledgments}

The authors are grateful to anonymous referees for a number of constructive comments which helped to improve the original manuscript. They would also like to appreciate S. M. M. Rasouli for useful discussions and comments on BD theory.

\appendix
\section{Lyra-Kretschmann Scalar}\label{appA}
The Lyra-Kretschmann scalar is defined as
\be
\mathcal{K}=K^{\mu}\,_{\nu\alpha\beta}\, K_{\mu}\,^{\nu\alpha\beta},
\ee
which for the spacetime given by  metric (\ref{metric}) can be computed as
\bea
\mathcal{K}=\f{12}{\psi^2}\Bigg[\alpha_1 H^4
+\alpha_2 H^2(\dot{H}+H^2)
+\alpha_3 (\dot{H}+H^2)^2\Bigg],
\eea
where
\bea
\alpha_1&=&\bigg[(s+1)^2(2a^3\psi^6\psi_{,a}\psi_{,aa}+
+a^4\psi^4\psi_{,a}^4-2a^4\psi^5\psi_{,a}^2\psi_{,aa}+7a^2\psi^2\psi_{,a}^2-2a^3\psi^5\psi_{,a}^3
+a^4\psi^6\psi_{,aa}^2)\nonumber\\
&+&2(2s+1)\big[a^2\psi^6\psi_{,a}^2-a^2\psi^4\psi_{,a}^2\big]+(s+1)\big[6a\psi^3\psi_{,a}+2a\psi^7\psi_{,a}-4a\psi^5\psi_{,a}\big]\nonumber\\
&+&(s^4+4s^3+6s^2+4s+1)a^4\psi_{,a}^4
+4(s^3+3s^2+3s+1)a^3\psi\psi_{,a}^3+\psi^8-2\psi^6+\psi^4\bigg],\\
\alpha_2&=&\bigg[(s+1)\Big(2s a^2\psi^6\psi_{,a}^2+2a\psi^7\psi_{,a}+2a^2\psi^7\psi_{,aa}\Big)+2(s+1)^2\Big(a^3\psi^6\psi_{,a}\psi_{,aa}-a^3\psi^5\psi_{,a}^3\Big)\bigg],\\
\alpha_3&=&\bigg[2(s+1)a\psi^7\psi_{,a}+(s+1)^2a^2\psi^6\psi_{,a}^2+\psi^8\bigg],
\eea
and $\psi^2=\phi$. Substituting for $H$ and $\phi$ from equations (\ref{H}) in the above expression, we finally get
\bea
\mathcal{K}=\frac{64 \pi ^2 \rho_{i_{m}} ^2 }{3 \beta^6 (s-1)^2 s^4}\bigg[&&-8 a^{\frac{2}{1-s^2}} \beta s^3(s-1)^2 (s+2)+4 a^{\frac{4}{1-s^2}} (s+1)^2 (s^4+2s^3-5s^2+4s-1)\nonumber\\
&&+\beta^2 s^2 \Big(5 s^4-14 s^2+16 s-3+6w(s^4-1)+9w^2(s^2-1)^2\Big)\bigg]a^{-\frac{6(1+w)(s^2-1)+8}{s^2-1}}.
\eea


\begin{thebibliography}{99}
\bibitem{HE} S. W. Hawking and G. F. R. Ellis, {\it The Large Scale Structure of Space-Time}, Cambridge University Press,
Cambridge (1973).
\bibitem{CCCP} R. Penrose, Riv. Nuovo Cimento \textbf{1} (1969) 252.
\bibitem{SF} R. Giamb\`{o}, Class. Quantum Grav \textbf{22} (2005) 2295;\\
S. Bhattacharya, R. Goswami and P. S. Joshi, Int. J. Mod. Phys. D \textbf{20} (2011) 1123;\\
S. Bhattacharya, Proceedings of JGRG19; arXiv:1107.4112 [gr-qc].
\bibitem{PF} T. Harada, Phys. Rev. D \textbf{58} (1998) 104015;\\
T. Harada and H. Maeda, Phys. Rev. D \textbf{63} (2001) 084022;\\
R. Goswami and P. S. Joshi, Class. Quantum Grav \textbf{19} (2002) 5229;\\
R. Giamb\`{o}, F. Giannoni, G. Magli and P. Piccione, Gen. Rel. Grav \textbf{36} (2004) 1279;\\
J. F. Villas da Rocha and A. Wang, Class. Qauntum Grav \textbf{17} (2000) 2589;\\
R. Giamb\`{o}, F. Giannoni, G. Magli and P. Piccione, Class. Quantum Grav \textbf{20} (2003) 4943.
\bibitem{IPF} P. Szekeres and V. Iyer, Phys. Rev. D \textbf{47} (1993)
4362;\\ S. Barve, T. P. Singh, and L. Witten, Gen. Rel. Grav. \textbf{32} (2000) 697;\\
A. A. Coley and B. O. J. Tupper, Phys. Rev. D \textbf{29} (1984) 2701;\\K. Lake, Phys. Rev. D \textbf{26} (1982) 518.
\bibitem{Khodam} A. H. Ziaie, K. Atazadeh and S. M. M. Rasouli, Gen. Rel. Grav \textbf{43} (2011) 2943.
\bibitem{Love} N. Bedjaoui, P. G. LeFloch, J. M. Mart\`{\i}n-Garc\`{\i}a and J. Novak, Class. Quantum Grav
\textbf{27} (2010) 245010.
\bibitem{GBG} H. Maeda, Phys. rev. D, \textbf{73} (2006) 104004.
\bibitem{shear} P. S. Joshi, N. Dadhich and R. Maartens, Phys. Rev. D {\bf 65} (2002) 101501;\\
P. S. Joshi, R. Goswami  and N. Dadhich, Phys. Rev. D {\bf 70} (2004) 087502;\\
S. M. C. V. Gon\c{c}alves, Phys. Rev. D {\bf 69}, (2004) 021502(R).
\bibitem{Hehl} F. W. Hehl, P. von der Heyde, G. David Kerlick and J. M. Nester, Rev. Mod. Phys {\bf 48} (1976) 393.
\bibitem{Lyra} G. Lyra, Math. Z. {\bf 54} (1951) 52;\\
E. Scheibe, Math. Z. {\bf 57} (1952) 65.
\bibitem{VDSabbata-Gas} V. de Sabbata and M. Gasperini, ``{\it Introduction to gravitation}\rq\rq{}, World Scientific Pub Co Inc; First Edition (July 1986).
\bibitem{BDT} C. Brans and R. H. Dicke, Phys. Rev {\bf 124} (1961) 925.
\bibitem{BDLyra} S. Hojman, M. Rosenbaum, M. P. Jr Ryan and L. C. Shepley, Phys. Rev. D {\bf 17} (1978) 3141;\\
S. Hojman, M. Rosenbaum and M. P. Jr Ryan, Phys. Rev. D {\bf 19} (1979) 430.
\bibitem{Soleng} H. H. Soleng, Class. Quantum Grav. {\bf 5} (1988) 1489.
\bibitem{KADUNN} K. A. Dunn,  J. Math. Phys. \textbf{15} (1974) 2229.
\bibitem{Sen-Ne} D. K. Sen, Z. Phys. \textbf{149} (1957) 311.
\bibitem{TOPO} F. Rahaman, S. Mal and M. Kalam, Astrophys. Space Sci. \textbf{319} (2009) 169;\\
F. Rahaman, P. Ghosh, Astrophys. Space Sci. \textbf{317} (2008) 127;\\
F. Rahaman, M. Kalam, R. Mondal, Astrophys. Space Sci. \textbf{305} (2006) 337;\\
A. Pradhan, I. Iotemshi, G. P. Singh,  Astrophys. Space Sci. \textbf{288} (2003) 315;\\
F. Rahaman, Int. J. Mod. Phys. D \textbf{9} (2000) 775;\\
F. Rahaman, Int. J. Mod. Phys. D \textbf{10} (2001) 579;\\
F. Rahaman, Astrophys. Space Sci. \textbf{283} (2003) 151;\\
F. Rahaman, S. Chakraborty and M. Kalam, Int. J. Mod. Phys. D \textbf{10} (2001) 735.
\bibitem{HDL} G. Mohanty, K. L. Mahanta, B. K. Bishi, Astrophys. Space Sci. \textbf{317} (2008) 283;\\
F. Rahaman, N. Begum, S. Das, Astrophys. Space Sci. \textbf{294} (2004) 219;\\
F. Rahaman, S. Chakraborty, S. Das, R. Mukherjee, M. Hossain, N. Begam, Astrophys. Space Sci. \textbf{288} (2003) 379;\\
G. Mohanty, K. L. Mahanta, Space Sci. \textbf{312} (2007) 301.
\bibitem{DEL} H. Hova, Journal of Geometry and Physics, \textbf{64} (2013) 146;\\
H. Zhi, M. Shi, X. Meng, L. Zhang, arXiv:1210.6431 [physics.gen-ph], DOI:10.1007/s10773-014-2151-4;\\
A. Pradhan, D. S. Chauhan, RAPC \textbf{8} (2009) 179;\\
K. S. Adhav, Int. J. Astron. Astrophys. \textbf{1} (2011) 204.
\bibitem{SPi} R. Casana, C. A. M. de Melo, B. M. Pimentel, Astrophys. Space Sci. \textbf{305} (2006) 125.
\bibitem{MDKP} R. Casana, C. A. M. de Melo, B. M. Pimentel, Class. Quant. Grav. \textbf{24} (2007) 723.
\bibitem{SVMF} R. Casana, C. A. M. de Melo, B. M. Pimentel, arXiv:hep-th/0501085.
\bibitem{SGGML} F. Rahaman, Astrophys. Space Sci. \textbf{283} (2003) 155.
\bibitem{LBB} F. Rahaman, B. C. Bhui, G. Bag, Astrophys. Space Sci. \textbf{295} (2005) 507;\\
G. S. Khadekar, A. Pradhan, K. Srivastava, arXiv:gr-qc/0508099.
\bibitem{2Singh} T. Singh and G. P. Singh, Fortschr. Phys. \textbf{41} (1993) 737.
\bibitem{LBH} F. Rahaman, A. Ghosh, M. Kalam, Nuovo Cim. B \textbf{121} (2006) 649.
\bibitem{Sen} D. K. Sen and J. R. Vanstone, J. Math. Phys {\bf 13} (1972) 990;\\
D. K. Sen and K. A. Dunn, J. Math. Phys, {\bf12} (1971) 578.
\bibitem{-3/2} V. Faraoni and N. Lanahan- Tremblay Phys. Rev. D {\bf 78} (2008) 064017;\\
N. Lanahan- Tremblay and V. Faraoni Class. Quantum Grav. {\bf 24} ( 2007) 5667;\\
T. P. Sotiriou and V. Faraoni,  Rev. Mod. Phys. {\bf 82} 451 (2010).
\bibitem{Frolov} V. P. Frolov and I. D. Novikov, {\it Black Hole Physics} , Copenhagen
$\&$ Edmonton, October (1997).
\bibitem{Hayward} S. A. Hayward, Phys. Rev. D \textbf{49} (1994) 6467;\\
S. A. Hayward, Phys. Rev. D \textbf{53} (1994) 1938.
\bibitem{Misner-Sharp} C. W. Misner and D. H. Sharp, Phys. Rev. \textbf{136}, (1964) B571;\\
D. Bak and S. J. Rey, Class. Quantum Grav. \textbf{17}, (2000) L83.
\bibitem{JAP} T. Koike, H. Onozawa, M. Siino, arXiv:gr-qc/9312012.
\bibitem{GMSMAG} H. Maeda, Phys. Rev. D \textbf{73} (2006) 104004;\\
Y. Gong and A. Wang, Phys. Rev. Lett. {\bf 99} (2007) 211301;\\
H. Maeda and M. Nozawa, Phys. Rev. D \textbf{77} (2008) 064031;\\
M. Nozawa and H. Maeda, Class. Quantum. Grav. \textbf{25} (2008) 055009;\\
S.-Feng Wu, B. Wang, G.-Hong Yang, Nucl. Phys. B \textbf{799} (2008) 330;\\
G. Kunstatter, H. Maeda and T. Taves, Class. Quantum Grav. \textbf{30} (2013) 065002.
\bibitem{cai} R. G. Cai, L. M. Cao, Y. P. Hu and N. Ohta, Phys. Rev. D {\bf 80}, (2009) 104016.
\bibitem{MEHR} S. M. M. Rasouli, M. Farhoudi and P. V. Moniz, Class. Quantum Grav. \textbf{31} (2014) 115002;\\N. Banerjee and S. Sen, Phys. Rev. D, \textbf{56} (1997) 1334;\\ V. Faraoni, Phys. Rev. D \textbf{59} (1999) 084021.
\bibitem{NEGP} P. Szekeres and V. Iyer Phys. Rev. D \textbf{47} (1993) 4362;\\ P. S. Joshi, R. Goswami, Class. Quantum Grav. \textbf{24} 2917 (2007);\\ R. Goswami and P. S. Joshi, arXiv:gr-qc/0504019;\\ R. Goswami and P. S. Joshi, C. Vaz and L. Witten, Phys. Rev. D \textbf{70} (2004) 084038;\\ P. S. Joshi, R. Goswami, arXiv:0711.0426 [gr-qc];\\ M. Patil, P. S. Joshi and D. Malafarina, Phys. Rev. D \textbf{83} (2011) 064007.
\bibitem{COOP} F. I. Cooperstock, S. Jhingan, P. S. Joshi, T. P. Singh, Class. Quantum Grav. \textbf{14} 2195 (1997).
\bibitem{MALAF} C. Bambi, D. Malafarina and L. Modesto, Eur. Phys. J. C \textbf{74} (2014) 2767;\\
C. Bambi, D. Malafarina, A. Marcianoand and L. Modesto, Phys. Lett. B \textbf{734} (2014) 27.
\bibitem{LQGSINA} M. Bojowald, R. Goswami, R. Maartens, P. Singh, Phys. Rev. Lett. \textbf{95} 091302 (2005);\\ C. Bambi, D. Malafarina and L. Modesto,  Phys. Rev. D \textbf{88} (2013) 044009.
\bibitem{MHJP} S. M. M. Rasouli, A. H. Ziaie, J. Marto, P. V. Moniz, Phys. Rev. D \textbf{89} 044028 (2014).
\bibitem{QGSA} G. M. Hossain, Class. Quant. Grav. \textbf{22} 2653 (2005);\\
Y. Tavakoli, J. Marto, and A. Dapor, Int. J. Mod. Phys. D \textbf{23} 1450061 (2014); arXiv:1303.6157;\\ P. Singh and A. Toporensky,Phys. Rev. D \textbf{69} 104008 (2004);\\ C. Bambi, D. Malafarina and L. Modesto, Eur. Phys. J. C \textbf{74} 2767 (2014); arXiv:1306.1668 [gr-qc];\\ A. Kreienbuehl, T. Pawlowski,\rq\rq{} Phys. Rev. D \textbf{88} (2013) 043504;\\ T. Cailleteau, A. Cardoso, K. Vandersloot and D. Wands, Phys. Rev. Lett. \textbf{101} (2008) 251302.
\bibitem{RoBe} T. A. Roman and P. G. Bergmann, Phys. Rev. D \textbf{28} 1265 (1983).
\bibitem{Wald-Iyer} R. M. Wald and V. Iyer, Phys. Rev. D \textbf{44} R3719 (1991);\\
Black Holes and Relativistic Stars, Edited by R. M. Wald, University of Chicago Press (1998).
\bibitem{JSITR} S. Jhingan, P. S. Joshi and T. P. Singh, Class. Quantum .Grav. \textbf{13} (1996) 3057;\\
R. Giambo and G. Magli, Differential Geometry and its Applications, \textbf{18} (2003) 285; arXiv:math-ph/0209059;\\ K. D. Patil, Int. J. Mod. Phys. D, \textbf{15} (2006) 251.
\bibitem{SHTE} S. L. Shapiro and S. A. Teukolsky, Phys. Rev. Lett. \textbf{66} 994 (1991);\\
S. L. Shapiro and S. A. Teukolsky, Am. Sci. \textbf{79} 330 (1991).
\bibitem{GI} R. Giambo, Class. Quantum Grav. \textbf{22} (2005) 2295.
\bibitem{Joshi}  P. S. Joshi, ``{\it Gravitational Collapse and Space-Time Singularities}'', Cambridge University Press (Cambridge, 2007).
\bibitem{TIP} F. J. Tipler, Phys. Lett. A \textbf{67} (1977) 8;\\
F. J. Tipler, C. J. S. Clarke and G. F. R. Ellis, in General
Relativity and Gravitation", edited by A. Held (Plenum,
New York, 1980), vol. 2, p. 97; C. J. S. Clarke, The Analysis of Space-time Singularities", Cambridge University
Press (Cambridge, 1993).
\bibitem{CL} C. J. S. Clarke Analysis of spacetime singularities (Cambridge University Press (1993)).
\bibitem{QPCNSC} U. Miyamoto, H. Maeda and T. Harada, Prog. Theor. Phys. \textbf{113} (2005) 513.
\bibitem{CLKR} C. J. S. Clarke and A. Krolak, J. Geo. Phys. \textbf{2} 127 (1986).
\bibitem{Vaidya} P. C. Vaidya, Proc. Indian Acad. Sci. A \textbf{33} (1951) 264; Reprinted, Gen. Rel. Grav. \textbf{31} (1999) 119; W. B. Bonnor, A. K. G. de Oliveira and N. O. Santos, \lq\lq{}Radiating spherical Collapse,\rq\rq{} Phys. Rep. \textbf{181} (1989) 269.
\bibitem{JOGLO} P. S. Joshi, {\it Global Aspects in Gravitation and Cosmology,} Clarendon, Oxford, (1993).
\bibitem{VAINS} S. G. Ghosh, Phys. Rev. D \textbf{62} (2000) 127505;\\
 A. Papapetrou, in\rq\rq{}{\it A random walk in general relativity,}\rq\rq{} ed. N. Dadhich, J. Krishna Rao, J. V. Narlikar and C. V. Vishveshwara, (Wiley Eastern, New Delhi 1985);\\ Y. Kuroda, Prog. Theor. Phys. \textbf{72} (1984) 63;\\ K. Lake, Phys. Rev. D \textbf{43} (1991) 1416;\\I. H. Dwivedi and P. S. Joshi, Class. Quantum Grav. \textbf{6} (1989) 1599; {\it ibid},  Class. Quantum Grav. \textbf{8} (1991) 1339;\\ S. M. Wagh, S. D. Maharaj, Gen. Rel. Grav. \textbf{31} (1999) 975;\\ A. Beesham, S. G. Ghosh,Int. J. Mod. Phys. D \textbf{12} (2003) 801;\\ T. Harko, K. S. Cheng, Phys. Lett.  A \textbf{266} (2000) 249;\\ S. G. Ghosh, N. Dadhich, Phys. Rev. D \textbf{64} (2001) 047501;\\ T. Harko, Phys. Rev. D \textbf{68} (2003) 064005.
\bibitem{LEM} J. P. S. Lemos, Phys. Rev. D \textbf{59} (1999) 044020.
\bibitem{GVMEX} A. Wang and Y. Wu, Gen. Rel. Grav., \textbf{31} (1999) 107;\\ S. D. Maharaj, G. Govender and M. Govender, Gen. Relativ. Gravit. \textbf{44} (2012) 1089.
\bibitem{IDJCC} W. Israel, Nuovo Cimento B, \textbf{44} (1996) 1; ibid, Nuovo
Cimento B 48 (1967) 463.
\bibitem{SURLAYER} K. Lake, in Vth Brazilian school of cosmology and gravitation edited by M. Novello (World Scientific, Singapore, 1987);\\
P. Musgrave and K. Lake, Class. Quantum Grav. \textbf{13} (1996) 1885; ibid, Class. Quantum Grav. \textbf{14} (1997) 1285.
\bibitem{HarkoST} T. Harko and K. S. Cheng, Phys. Lett. A \textbf{266} (2000) 249.
\bibitem{STFL} E. N. Glass and J. P. Krisch, Phys. Rev. D \textbf{57} (1998) 5945; Class. Quant. Grav. \textbf{16} (1999) 1175;\\S. G. Ghosh, Int. J. Mod. Phys. A \textbf{23} (2008) 4245.
\bibitem{Lindquist R.W} R. W. Lindquist, R. A. Schwartz, and C. W. Misner, Phys. Rev. B \textbf{137} 1364
(1965).
\bibitem{FAYFL} F. Fayos and R. Torres, Class. Quantum Grav. \textbf{28} 105004 (2011).
\bibitem{GIAMKH} R. Giambo, J. Math. Phys., \textbf{50} (2009) 012501.
\bibitem{JD} I. J. Dwivedi and P. S. Joshi, Class. Quantum Grav. \textbf{6} 1599 (1989).
\bibitem{CCC} A. Kr\'{o}lak, Prog. Theor. Phys. Suppl. \textbf{136} (1999) 45;\\
P. R. Brady, I. G. Moss and R. C. Myers, Phys. Rev. Lett. \textbf{80} (1998) 3432;\\
R. M. Wald, \lq\lq{}Gravitational Collapse and Cosmic Censorship\rq\rq{}, [arXiv:gr-qc/9710068].
\bibitem{RECJO} P. S. Joshi, D. Malafarina,  Int. J. Mod. Phys. D, \textbf{20} 2641 (2011).
\bibitem{NSVSBH} P. S. Joshi, D. Malafarina, and R. Narayan, Class. Quantum Grav. \textbf{28} 235018 (2011).
\bibitem{VirEllis} K. S. Virbhadra, D. Narasimha, and S. M. Chitre, Astron. Astrophys. \textbf{337} 1 (1998);\\
K. S. Virbhadra, and C. R. Keeton, Phys. Rev. D \textbf{77} 124014 (2008);\\
K. S. Virbhadra, and G. F. R. Ellis, Phys. Rev. D \textbf{65} 103004 (2002); Phys. Rev. D \textbf{62} 084003 (2000);\\
C-M Claudel, K. S. Virbhadra, and G. F. R. Ellis, J. Math. Phys. \textbf{42} 818 (2001).
\bibitem{TPAG} V. De Sabbata and M. Gasperini, Phys. Rev. D \textbf{23} 2116 (1981);\\
A. Saa, Gen. Rel. and Grav. \textbf{29} 205 (1997); Mod. Phys. Lett. A \textbf{8} 2565 (1993); Mod. Phys. Lett. A \textbf{9} 971 (1994); Class. Quant. Grav. \textbf{12} L85 (1995); J. Geom. and Phys. \textbf{15} 102 (1995).
\bibitem{CAR} S. M. Carrol and G. B. Field, Phys. Rev. D \textbf{50} 3867 (1994).
\bibitem{EXTOR} R. T. Hammond Phys. Rev. D \textbf{52} 6918 (1995).
\bibitem{LOWTOR} V. G. Bagrov, I. L. Buchbinder, I. L. Shapiro, arXiv:hep-th/9406122;\\
P. Singh and L. H. Ryder, Class. Quantum Grav. \textbf{14} (1997) 3513.
\bibitem{SHAP} A. S. Belyaev and I. L. Shapiro, Nuclear Physics  B  \textbf{543} 20 (1999);\\
I. L. Shapiro, Phys. Rep. \textbf{357} 113 (2002);\\
A. S. Belyaev and I. L. Shapiro, Phys. Lett. B, \textbf{425} 246 (1998).
\bibitem{EARTOR} G. G. A. Bauerle, Chr. J. Haneveld Physica A, \textbf{121} 541 (1983);\\M. Gasperini, Phys. Rev. Lett. \textbf{56} 2873 (1986); Gen. Rel. Grav. \textbf{30} 1703 (1998);\\
V. De Sabbata, Nuovo Cimento A \textbf{107} 363 (1994);\\ G. de Berredo-Peixoto, E. A. de Freitas, Class. Quant. Grav. \textbf{26} 175015 (2009).
\bibitem{OSD} J. R. Oppenheimer and H. Snyder, Phys. Rev. \textbf{56} (1939) 455;\\
 S. Datt, Zs. f. Phys. \textbf{108} (1938) 314.
 \bibitem{LTB} A. Ori and T. Piran, Phys. Rev. Lett. \textbf{59} 2137 (1987);\\
F. J. Tipler, Phys. Lett. A \textbf{64} (1987) 8;\\
F. C. Mena, R. Tavakol and P. S. Joshi, Phys. Rev. D \textbf{62} (2000) 044001;\\
P. S. Joshi and I. H. Dwivedi, Commun. Math. Phys. \textbf{166} (1994) 117; Class. Quantum
Grav. \textbf{16} (1999) 41;\\
D. Christodoulou, Commun. Math. Phys. \textbf{93} (1984) 171;\\
R. P. A. C. Newman, Class. Quantum. Grav. \textbf{3} (1986) 527;\\
P. S. Joshi, N. Dadhich and R. Maartens, Phys. Rev. D \textbf{65} (2002) 101501(R);\\
A. Banerjee, U. Debnath and S. Chakraborty, Int. J. Mod. Phys. D \textbf{12} (2003) 1255;\\
U. Debnath and S. Chakraborty, Gen. Rel. Grav. \textbf{36} (2004) 1243;\\
U. Debnath and S. Chakraborty, Gen. Rel. Grav. \textbf{37} (2005) 225;\\
S, Jhingan and P. S. Joshi, arXiv:gr-qc/9701016.

\bibitem{GSPGS} S. H. Ghate, R. V. Saraykar and K. D. Patil, Pramana J. Phys. \textbf{53} (1999) 253;\\
P. S. Joshi, D. Malafarina and R. V. Saraykar, Int. J. Mod. Phys. D \textbf{21} (2012) 1250066;\\ R. Goswami, P. S. Joshi and D. Malafarina, arXiv:1202.6218 [gr-qc];\\ N. Ortiz, AIP Conf. Proc. \textbf{1473} (2011) 49;\\ R. V. Saraykar and P. S. Joshi, arXiv:1207.3469 [gr-qc].
\bibitem{GTG} M. Blagojevic and ‎F. W. Hehl, \lq\lq{}Gauge Theories of Gravitation: A Reader with Commentaries,\rq\rq{} Imperial College Press (London) 2013; F. Gronwald, F. W. Hehl, \lq\lq{}On the Gauge Aspects of Gravity,\rq\rq{} arXiv:gr-qc/9602013; \\G. de Berredo Peixoto, J. A. Helayel-Neto and I. L. Shapiro, JHEP 0002 (2000) 003; \\F. W. Hehl, J. D. McCrea, E. W. Mielke and Y. Ne\rq{}eman, Phys. Rep. \textbf{258}, (1995) 1;\\ V. De Sabbata and M. Gasperini, Phys. Lett. A \textbf{77} (1980) 300; Phys. Lett. A \textbf{83} (1981) 115.

\bibitem{REQFT}  I. L. Buchbinder,  I. L.  Shapiro,  Phys.  Lett.  B  \textbf{151} (1985) 263; Class. Quantum Grav. \textbf{7} (1990) 1197;\\ I. L. Buchbinder, S. D. Odintsov and I. L. Shapiro, \lq\lq{}Effective action in quantum gravity,\rq\rq{}  (IOP,  Bristol,  1992);\\ L. H. Ryder and I. L. Shapiro, Phys. Lett. A \textbf{247} (1998) 21.
\bibitem{Kostel} V. A. Kostelecky, N. Russell and J. D. Tasson, Phys. Rev. Lett. \textbf{100} (2008) 111102.
\bibitem{HDE} C. Lammerzahl, Phys. Lett. A \textbf{228} (1997) 223.
\bibitem{HE4} R. Lehnert, W. M. Snow and H.Yan, Phy. Lett. B \textbf{730} (2014) 353.
\bibitem{ABWZ} J. Alsing, E. Berti, C. M. Will and H. Zaglauer, Phys. Rev. D \textbf{85} (2012) 064041.
\bibitem{PS} D. Psaltis, Living Rev. Relativity, \textbf{11} (2008) 9.
\bibitem{CMWill} C. M. Will, Living Rev. Relativity, \textbf{9} (2006) 3.
\bibitem{VF} V. Faraoni, \lq\lq{}Cosmology in Scalar-Tensor Gravity,\rq\rq{} Kluwer Academic Publishers (2004).
\bibitem{PRD-55-1906} J. D. Barrow and P. Parsons, Phys. Rev. D \textbf{55} (1997) 1906.
\bibitem{TDBBN} T. Damour, and B. Pichon, Phys. Rev. D \textbf{59} (1999) 123502.
\bibitem{ACBBN} A. Coc, K. A. Olive, J. P. Uzan and E. Vangioni, Phys. Rev. D \textbf{73} (2006) 083525.
\bibitem{TBS} T. Clifton, J. D. Barrow and R. J. Scherrer, Phys. Rev. D \textbf{71} 123526 (2005).
\bibitem{UZAN} J.-P. Uzan, Rev.  Mod. Phys, \textbf{75} 403 (2003).
\bibitem{Scheel} M. A. Scheel, S. L. Shapiro, and S. Teukolsky, Phys. Rev. D \textbf{51} (1995) 4236.
\end{thebibliography}
\end{document}